\documentclass[10pt]{iopart}

\usepackage{color}
\usepackage{graphicx}% Include figure files
\usepackage{braket}

\newcommand{\jeff}{\ensuremath{{j_\mathrm{eff}}}}
\newcommand{\Dtri}{\ensuremath{\Delta_\mathrm{tri}}}

\newcommand{\Ueff}{\ensuremath{U_\mathrm{eff}}}

\newcommand{\egp}{{\ensuremath{e_\mathrm{g}^\prime}}}

\newcommand{\ttg}{\ensuremath{{t_\mathrm{2g}}}}

\def\Ueff{U_\mathrm{eff}}

\def\Asia{A_\mathrm{sia}}
\def\ADM{A_\mathrm{DM}}

\begin{document}

\title{First-principles studies of spin-orbital physics in pyrochlore oxides}

\author{Hiroshi Shinaoka}
\address{Department of Physics, Saitama University, 338-8570, Japan}
\ead{shinaoka@mail.saitama-u.ac.jp}

\author{Yukitoshi Motome}
\address{Department of Applied Physics, University of Tokyo, Tokyo 113-8656, Japan}

\author{Takashi Miyake}
\address{Research Center for Computational Design of Advanced Functional Materials (CD-FMat), National Institute of Advanced Industrial Science and Technology (AIST), Umezono, Tsukuba 305-8568, Japan}

\author{Shoji Ishibashi}
\address{Research Center for Computational Design of Advanced Functional Materials (CD-FMat), National Institute of Advanced Industrial Science and Technology (AIST), Umezono, Tsukuba 305-8568, Japan}

\author{Philipp Werner}
\address{Department of Physics, University of Fribourg, 1700 Fribourg, Switzerland}

\vspace{10pt}
\begin{indented}
\item[]March 2017
\end{indented}

%\maketitle

\begin{abstract}
The pyrochlore oxides $A_2B_2$O$_7$ exhibit a complex interplay between geometrical frustration, electronic correlations, and spin-orbit coupling, due to the lattice structure and active charge, spin, and orbital degrees of freedom. 
Understanding the properties of these materials is a theoretical challenge, because their intricate nature depends on material-specific details and quantum many-body effects.
Here we review our recent studies based on first-principles calculations and quantum many-body theories for 4$d$ and 5$d$ pyrochlore oxides with $B$=Mo, Os, and Ir.
In these studies, the spin-orbit coupling and local electron correlations are treated within the LDA+$U$ and LDA+dynamical mean-field theory formalisms. We also discuss the technical aspects of these calculations.
\end{abstract}

%\maketitle
\ioptwocol

\section{Introduction}

Electrons in solids possess not only charge but also spin and orbital degrees of freedom. 
Although the charge degree of freedom plays a dominant role in conventional metals and semiconductors, the situation is drastically altered in systems with strong repulsive Coulomb interactions and relativistic spin-orbit coupling (SOC). 
In such systems, spin and orbital are also activated and the different degrees of freedom are entangled with each other, leading to a variety of intriguing cooperative phenomena, such as metal-insulator transitions~\cite{Imada:1998zz}, unconventional superconductivity~\cite{Norman196,doi:10.1080/00018732.2017.1331615}, and cross correlation phenomena like the magnetoelectric effect~\cite{0022-3727-38-8-R01,Fiebig2016}.

The pyrochlore oxides, with chemical formula $A_2B_2$O$_7$, provide a suitable playground to study phenomena arising from these multiple degrees of freedom. 
These compounds exhibit many fascinating properties which depend on the chemical constituents $A$ and $B$, and on  external fields like magnetic fields or pressure~\cite{Subramanian:1983ia,Gardner:2010fu}.
The $A$ and $B$ sites are typically occupied by rare-earth and transition metal atoms, respectively.
The $B$-site transition metal ions often play a crucial role through the multiple active degrees of freedom. 
In the case of $4d$ or $5d$ transition metals, such as Ru, Mo, Os, Re, and Ir, the electron interactions and the SOC can compete with each other. By changing the $A$-site ions, one can vary the filling of the $4d$ and $5d$ shells. 
This allows a systematic investigation of the keen competition between spin, charge, and orbital degrees of freedom. % in a comprehensive way.   

Recently, first-principles calculations 
have attracted growing interest as a powerful tool to study such complicated systems.
While the band structure calculations based on the density functional theory (DFT)~\cite{Hohenberg:1964zz,Kohn:1965ui}  
have some difficulties in the treatment of strong electron correlations in transition-metal oxides,  
substantial efforts have been devoted to the development of new frameworks combining the first-principles calculations with numerical simulations based on quantum many-body theory (for a review, refer to Refs.~\cite{Imada:2010epb,Held:2001cv,Kotliar:2006fl}).
To understand the diverse phenomena in the $4d$ and $5d$ pyrochlore oxides, it is obviously desired to capture the material-specific details and to take into account strong electron interactions and the SOC.
This is a challenging task, but several important aspects have been unveiled by the application of the recently developed combined frameworks.

In this article, we review such theoretical efforts aimed at explaining the intriguing phenomena in $4d$ and $5d$ pyrochlore oxides, with a focus on our studies based on first-principles calculations and quantum many-body theories. 
We discuss how the multiple degrees of freedom of the $4d$ and $5d$ electrons bring about the various properties observed in these systems.
We also elaborate on some technical aspects of the theoretical frameworks used in these studies. % of the combined methods between first-principles calculations and quantum many-body theories.

The review is structured as follows.
Section~2 is devoted to a short description of the crystal structure.
In section~3, after describing the general local electronic structure of 4$d$ and 5$d$ pyrochlores (section 3.1),
we briefly review spin-orbital physics in Mo, Os, and Ir pyrochlores.
More detailed discussions on Mo, Os and Ir pyrochlores are provided in sections 4, 5, and 6, respectively.
In section 7 we describe the theoretical framework used in the studies reviewed in this article.
A summary and future perspectives are given in section 8.

\section{Crystal structure}
In this section, we provide a short description of the crystal structure of the cubic pyrochlore oxides with the general formula $A_2$$B_2$O$_7$,
where $A$ is a rare earth and $B$ is a transition metal.
For different elements $A$ and $B$,
the cubic pyrochlore oxides exhibit a variety of physical properties. 
There exist several general reviews on the cubic pyrochlores~\cite{Subramanian:1983ia, Gardner:2010fu}.
In particular, the magnetic properties of pyrochlore oxides were discussed in detail in Ref.~\cite{Gardner:2010fu}.
As mentioned in the introduction,
the pyrochlore oxides have attracted much attention in condensed matter physics since they exhibit intriguing phenomena ranging from metal-insulator transitions to geometrically frustrated magnetism and spin-orbital physics. In this section, we will explain how spin-orbital and multi-orbital physics emerges in the pyrochlores by considering 
their crystal structures.

\begin{figure}
	\includegraphics[width=.5\textwidth,clip]{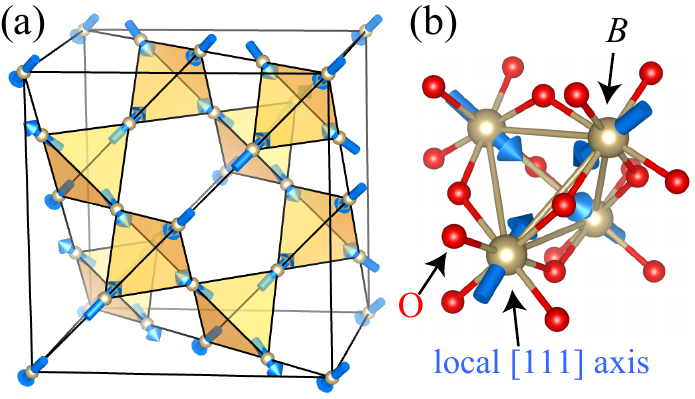}
	\caption{
		%Crystal structure of $A_2$$B_2$O$_6$O$^\prime$ [(a)] and oxygen atoms forming octahedra around the $B$ atoms [(b)].
		Crystal structure of (a) the cubic pyrochlore oxides $A_2B_2$O$_7$ and (b) a $B$ tetrahedron with surrounding oxygen atoms.
		In (a), we only show the $B$ atoms.
		The arrows indicate the local [111] axes.
	}
	\label{fig:crystal-struc}
\end{figure}

The cubic pyrochlore oxides crystallize in the space group $Fd\bar{3}m$ (No. 227).
Both the $A$ and $B$ sites form a corner-sharing network of tetrahedra.
In this review, we focus on the network of $B$ sites, where transition metal ions exhibit multi-orbital physics
\footnote{Although our focus is on the $B$ sites, intriguing physics arising from exchange interactions between $f$ moments on the $A$ sites and $d$ electrons on the $B$ sites has been extensively studied~\cite{Huang:2013kq,Chen:2012eka,Yao:2018fi}.}.
This network forms a pyrochlore lattice, 
as illustrated in Fig.~\ref{fig:crystal-struc}(a).
There are two kinds of tetrahedra formed by the $B$ sites, one of which is upside-down with respect to the other one.
We note that a similar pyrochlore lattice appears in the spinels $AB_2$O$_4$ as the network of the $B$ sites.

The pyrochlore lattice is a face-centered-cubic (fcc) lattice of tetrahedra, whose primitive unit cell contains four $B$ atoms.
It can be viewed as an alternating stacking of triangular and kagome lattices along the [111] direction.
In many pyrochlore oxides and related compounds, this peculiar type of corner-sharing network of transition metal ions gives rise to geometrically frustrated magnetism~\cite{Gardner:2010fu}.

Let us focus on the role of the oxygen atoms in the crystal structure,
since they are the source of rich orbital physics.
To be precise, there are two crystallographically inequivalent positions of oxygen atoms.
This is explicitly shown by writing the chemical formula $A_2$$B_2$O$_6$O$^\prime$.
In technical terms of crystallography,
the former O sites correspond to Wyckoff position $48f$, and the O$^\prime$ site to $8b$.
The $B$ sites are octahedrally coordinated, being surrounded by an octahedron of six oxygen atoms in the Wyckoff position $48f$.
For more details on the crystal structure, 
we refer to the review article by Subramanian \textit{et al.}~\cite{Subramanian:1983ia}.

The octahedral coordination of the $B$ sites plays an important role in the spin-orbital physics of the cubic pyrochlore.
There is only one adjustable positional parameter for the O site in $48f$.
This positional parameter is denoted by $x(\mathrm{O}_1)$ or simply by $x$ in the literature~\footnote{There are several possible choices of origin for describing the crystal structure. Here, we use the convention where the $B$ cation is chosen as the origin.}. 
We have a perfect octahedron for $x=0.3125$. 
For $x > 0.3125$, the oxygen octahedra are \textit{compressed} along the local [111] axes illustrated in Fig.~\ref{fig:crystal-struc}(b).
The value of $x$ is typically in the range of 0.320--0.345 (see Sec. II A 2 of Ref.~\cite{Gardner:2010fu}).
Note that the local [111] axis is different for the four $B$ sites in a unit cell.
This local trigonal distortion not only lifts the degeneracy of the $t_\mathrm{2g}$ orbitals but also changes the $B$--O--$B$ angle.
These changes in the crystal field and bond angle may affect the electronic structure in nontrivial ways.
First-principles calculations are helpful to provide insights into the electronic properties of these compounds.

\section{Spin-orbital physics}
\subsection{Effects of crystal field and spin-orbit coupling}
\begin{figure}
	\centering
	\includegraphics[width=.5\textwidth,clip]{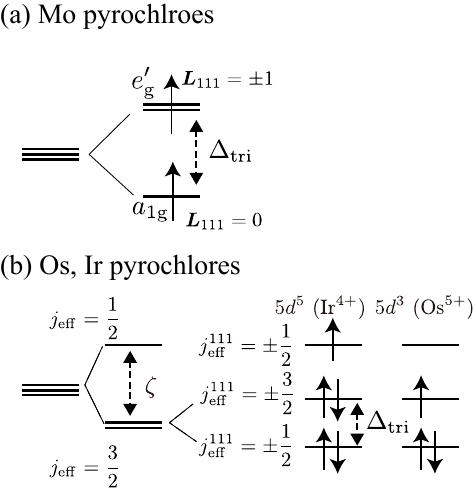}
	\caption{Crystal-field splitting of the $\ttg$ orbitals for Mo pyrochlores [(a)] and Os, Ir pyrochlores [(b)].
		The filling of the $d$ shell is indicated by arrows for each compound.
	}
	\label{fig:CF}
\end{figure}

Figure~\ref{fig:CF} illustrates the typical crystal field splitting of 
4$d$ and $5d$ $t_\mathrm{2g}$ orbitals,
which induces rich spin-orbital physics in pyrochlore compounds, as we are going to discuss in this review.
%In $5d$ systems, the spin-orbit coupling $\zeta$ may be larger than the trigonal crystal field $\Dtri$.
%For 4$d$ systems, we first discuss the effects of $\Dtri$ since $\zeta$ is relatively small.
In the pyrochlore oxides with $4d$ electrons, the effect of the trigonal crystal field $\Dtri$ is relatively large compared to the SOC $\zeta$.
Under $\Dtri$, the $t_\mathrm{2g}$ orbitals split into doubly degenerate $e_\mathrm{g}^\prime$ orbitals and a non-degenerate $a_\mathrm{1g}$ orbital, as shown in Fig.~\ref{fig:CF}(a).
The $a_\mathrm{1g}$ orbital is elongated along the local [111] axis on each B atom [see Fig.~\ref{fig:crystal-struc}(b)].
The $a_\mathrm{1g}$ and $\egp$ orbitals have quantized orbital momenta along the local [111] axis.
The splitting of the $a_\mathrm{1g}$ and $\egp$ orbitals for 4$d$ and $5d$ pyrochlores is typically a fraction of an eV, which may be comparable to the Hund's coupling in magnitude.
In 4$d$ Mo pyrochores such as Y$_2$Mo$_2$O$_7$, each Mo$^{4+}$ ion is in a $4d^2$ electron configuration.
The Hund's coupling aligns the two spins parallel as illustrated in Fig.~\ref{fig:CF}(a), leaving the orbital degeneracy of the $\egp$ orbitals.
In later sections of this review,
we will see that this active orbital degree of freedom cooperates with
the SOC, which results in non-trivial spin-orbital physics.

The one-particle crystal-field levels for $5d$ pyrochlores may be better understood by first considering the effect of the SOC $\zeta$, as it is usually larger than $\Dtri$.
%Under $\zeta$, 
As shown in Fig.~\ref{fig:CF}(b), the $t_\mathrm{2g}$ orbitals split into a doublet ($j_\mathrm{eff}=1/2$) and a quartet ($j_\mathrm{eff}=3/2$) by $\zeta$.
Here, the effective total angular momentum $\hat{j}_\mathrm{eff} \equiv \hat{L} - \hat{S}$ is quantized
\cite{Kim:2008gi,Kim09Science,SuganoCF}.
The trigonal crystal field $\Dtri$
splits this quartet state further into two doublets, as shown in Fig.~\ref{fig:CF}(b).
In this case, $\hat{j}_\mathrm{eff}$ is no longer a good quantum number, but its projection onto  the local [111] axis (referred as $j_\mathrm{eff}^{111}$ later) still remains quantized.
For convenience, we denote these states as ``$j_\mathrm{eff}=1/2$" and ``$j_\mathrm{eff}=3/2$" states.

%In this review, we discuss the properties of pyrochlore iridates with a 5$d^5$ electron configuration.
Pyrochlore iridates are characterized by a 5$d^5$ electron configuration.
In these compounds, based on the simple one-particle picture mentioned above,
the $j_\mathrm{eff}=3/2$ quartet is fully filled, leaving the $\jeff$=1/2 state half filled.
As a consequence, the properties of the pyrochlore iridates are often discussed based on an effective single-band model for the $\jeff$=1/2 state.
On the other hand, such a simple single-band picture cannot be applied to the Os pyrochlores with a $5d^3$ configuration, as discussed in section~\ref{sec:osmate}.

In the above arguments, we assumed the $t_\mathrm{2g}$ orbitals to be described by Wannier functions hybridized with oxygen orbitals.
In some of the literature~\cite{Hozoi:2014ft},
the local electronic structure was discussed in terms of 
atomic-orbital-like $t_\mathrm{2g}$ orbitals, which causes some confusion about the definition of $\Dtri$.
In the present convention, the effects of $B$-$O$ hybridizations are included in $\Dtri$ in addition to the effects of longer-range ligand fields from the $A$ sites.
The same convention of $\Dtri$ as in this review is used in the analysis in the supplemental material of Ref.~\cite{Zhang:2017ib}.
The above-mentioned effects are however not included in the case of the atomic-orbital-like basis.

\begin{figure}
	\includegraphics[width=.5\textwidth,clip]{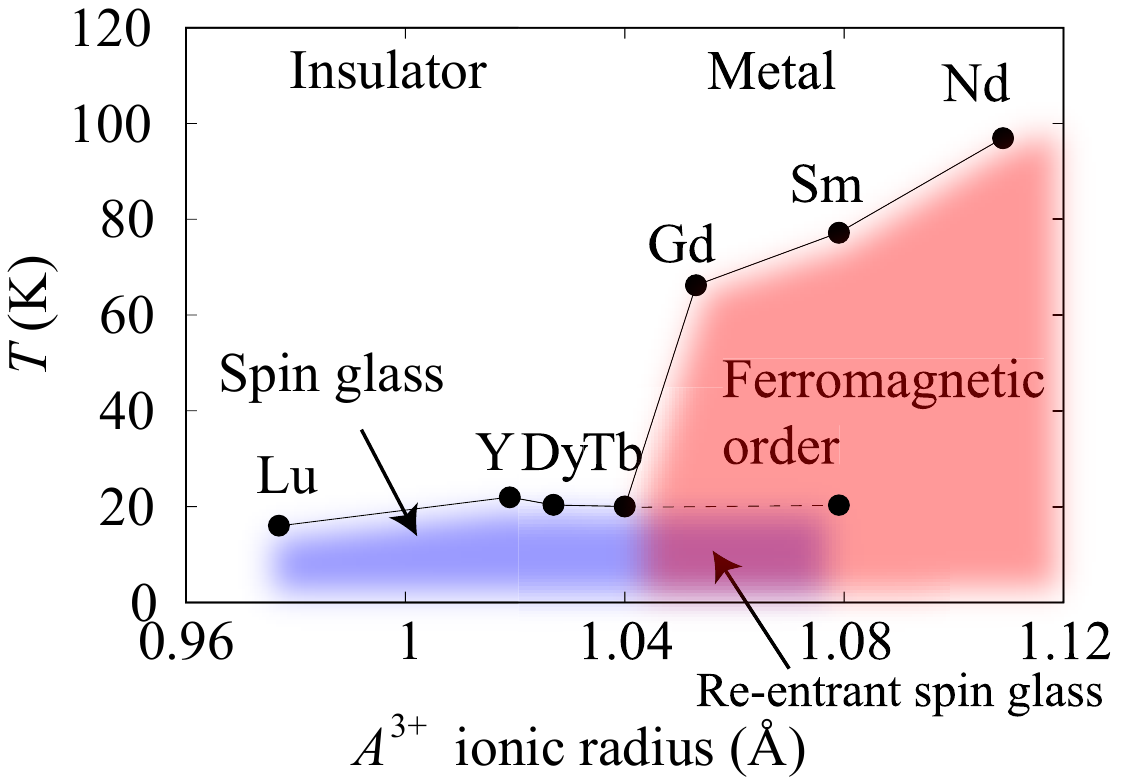}
	\caption{Experimental phase diagram of the Mo pyrochlores $A_2$Mo$_2$O$_7$. The experimental data are taken from Refs.~\cite{Sato:1986fm, 2000PhRvL..84.1998K, Gingras:1997fa,Prando:2014gn}.	

}
	\label{fig:Mo-pd}
\end{figure}

\subsection{Mo pyrochlores $A_2$Mo$_2$O$_7$}
Let us discuss several key properties of each compound in more depth.
The Mo pyrochlores $A_2$Mo$_2$O$_7$ host various interesting phenomena like metal-insulator transitions.\footnote{Recent experimental results are reviewed in section III. B of Ref.~\cite{Gardner:2010fu}.}
Figure~\ref{fig:Mo-pd} shows an experimental phase diagram with respect to the $A$-site ionic radius and temperature.
%As seen in Fig.~\ref{fig:Mo-pd}, compounds 
Compounds with relatively large $A$-site ionic radii ($A$=Nd, Sm) are metallic and exhibit a ferromagnetic transition at low $T$.
In a previous LDA+$U$ study~\cite{Solovyev:2003iv}
it was proposed that the ferromagnetism is due to the double exchange mechanism.
The essential idea is that the $a_\mathrm{1g}$ band is narrower than the $\egp$ band, %and can 
so that the corresponding electrons 
act as localized spins.
The relatively broader $\egp$ band can host itinerant electrons, leading to the double-exchange mechanism.
For smaller $A$-site ionic radii,
the ferromagnetic transition temperature decreases monotonically, and at the same time, the temperature dependence of the DC resistivity becomes insulating in the high-temperature paramagnetic phase.
Some compounds with small ionic radii ($A$=Y \textit{et al.}) show insulating transport properties at room temperature.

One interesting phenomenon, which is commonly observed in these insulating compounds, is spin-glass behavior at low $T$.
A spin glass is a magnetic state in which spins are frozen in random directions without spatial periodicity.
Conventional spin glasses have been observed in dilute magnetic systems where the magnetic exchange couplings alternate randomly in sign~\cite{Binder:1986zz}.
However, it is believed that Mo pyrochlores are not dirty enough to host such a conventional spin glass.\footnote{Refer to Ref.~\cite{Booth2000} and discussions in section III.B.1 of Ref.~\cite{Gardner:2010fu}.}
Several experimental results indicated that there is a re-entrant spin-glass phase on the metallic side~\cite{Iguchi:2009iva,Hanasaki:2007fwa,Prando:2014gn}.
A fundamental question thus is what causes this apparently unconventional spin-glass state. The geometrical frustration is a possible key element. 
%It is commonly believed that geometrical frustration plays an important role in the appearance of the spin-glass state.
To be more specific, geometrical frustration induces  competition between different magnetic states, which may somehow lead to the freezing of spins in random directions.
Thus, many theoretical studies have focused on the roles of the spin degree of freedom~\cite{Saunders:2007iz,Andreanov:2010dh,Shinaoka:2011kn, Shinaoka:2014kh}.

On the other hand, a previous first-principles study~\cite{Solovyev:2003iv} already revealed the importance of the orbital degree of freedom.  
Solovyev performed first-principles calculations (including the SOC) and studied magnetic ordering and metal-insulator transitions in these compounds.
It was revealed that in a magnetically ordered insulating state both the spin and orbital moments are active are oriented in an anti-parallel alignment.
This clearly indicates the importance of the SOC and the resultant spin-orbital composite degree of freedom. 

The role of the orbital degree of freedom in the spin-glass phenomena was not studied in more depth until the recent report of the first neutron scattering experimental results for single crystals of Y$_2$Mo$_2$O$_7$~\cite{Silverstein:2014bka}. 
A surprising observation was that the measured spin structure factor is not consistent with the conventional picture of geometrical frustration based on antiferromagnetic spin models.
This experimental finding motivated some of the authors to investigate the role of orbitals and the SOC in more depth~\cite{PhysRevB.88.174422}. 
In section 4, we describe how the orbital degree of freedom and the SOC give rise to a non-trivial magnetic frustration in the Mo pyrochlores.

\begin{figure}
	\includegraphics[width=.5\textwidth,clip]{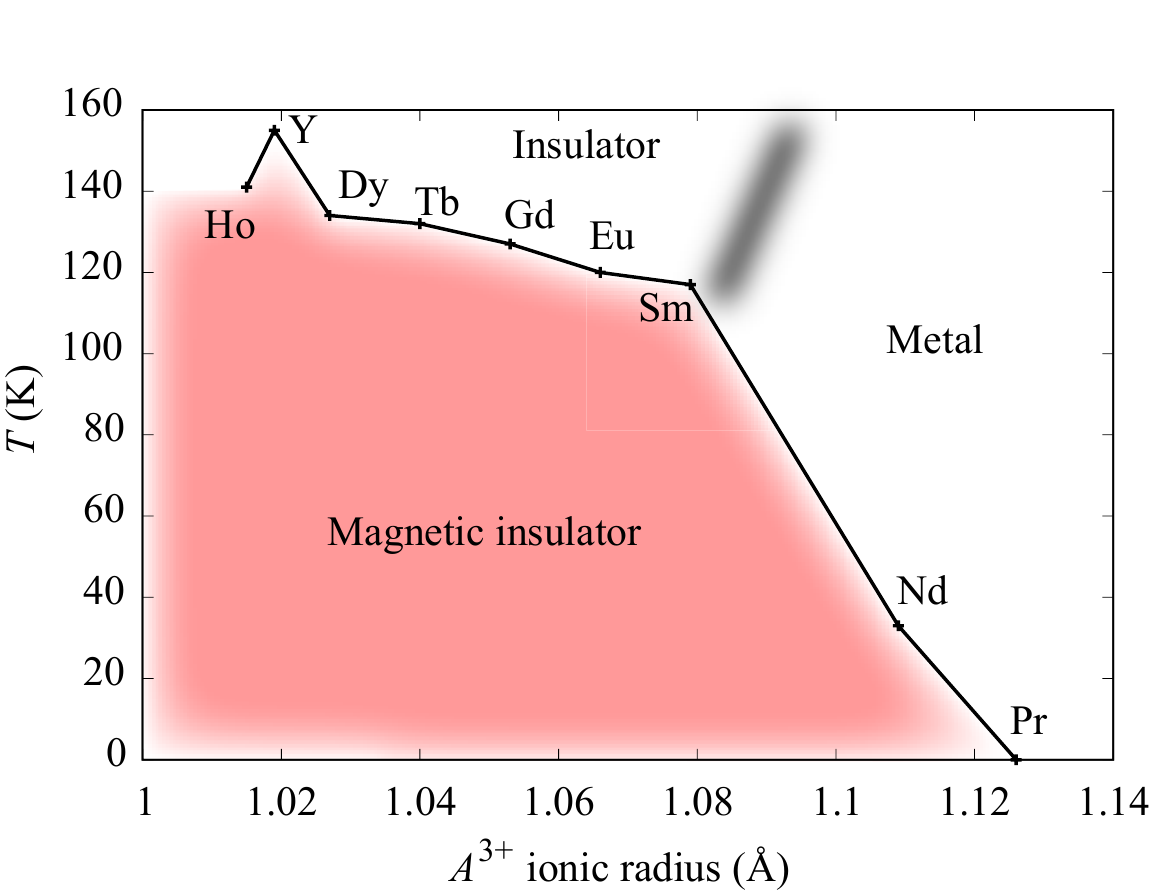}
	\caption{Experimental phase diagram of the pyrochlore iridates $A_2$Ir$_2$O$_7$.
		The experimental data are from Refs.~\cite{Matsuhira:2011ku,Ueda:2016dp}.}
	\label{fig:iridate-pd-exp}
\end{figure}

\subsection{Cd$_2$Os$_2$O$_7$}
In the Os and Ir pyrochlores, different numbers of electrons occupy the 5$d$ $t_\mathrm{2g}$ orbitals, although both compounds are known to exhibit metal-insulator transitions as $T$ is changed. 
The electronic properties of Cd$_2$Os$_2$O$_7$ with a 5$t_\mathrm{2g}^3$ configuration were reported for the first time in 1974~\cite{Sleight1974}.
It was revealed that the temperature dependence of the DC resistivity becomes insulating below $\sim$227 K %as well as the existence a cusp in 
and that a cusp appears in 
the magnetic susceptibility at the same $T$~\cite{Sleight1974}.
These indicate that the metal-insulator transition is associated with some magnetic ordering.
There have been many experimental and theoretical studies on the origin of the transition since the first report~\cite{Mandrus:2001ia,Padilla:2002ep,Harima:2002km,Singh:2002gt,Koda:2007ji,Matsuda:2011bk}.
However, Cadmium has a high neutron absorption cross-section, which prevents the use of neutron scattering techniques for determining the magnetic structure.
In 2012, Yamaura \textit{et al.} reported the results of X-ray diffraction and Raman scattering experiments~\cite{Yamaura:2012ey}.
They revealed the existence of the all--in--all--out magnetic order below the transition temperature. 
In section 5, we investigate the magnetic and electronic properties of the low-$T$ phase using first-principles calculations~\cite{Shinaoka:2012ja}.

\subsection{$A_2$Ir$_2$O$_7$ ($A$=rare earth) }
Figure~\ref{fig:iridate-pd-exp} shows a phase diagram summarizing recent experimental data of $A_2$Ir$_2$O$_7$ with a $5d^5$ configuration with respect to the $A$-site ionic radii and temperature.
In the early 2000s,
the existence of metal-insulator and antiferromagnetic transitions
was experimentally revealed~\cite{Yanagishima:2001ej,Taira:2001uz}. 
However, natural isotopes of iridium are also known as strong neutron absorbers, which has prevented the experimental investigation of  
the magnetic structure of the low-$T$ phases.

The low-$T$ properties of pyrochlore iridates have attracted much attention since a Weyl semimetal with nontrivial magneto-electric properties was proposed for the ground states of some compounds in this series in 2011~\cite{Wan:2011hi}. 
On the basis of LDA+$U$ calculations it has been argued that (i) some compounds are magnetically ordered in the so-called all--in--all--out structure, and
(ii) there exists a Weyl semimetal phase with the all--in--all--out order in the vicinity of an insulating phase. 
Indeed, the all--in--all--out magnetic order has been confirmed
in some pyrochlore iridates later by experiments ($A$=Nd, Eu)~\cite{Tomiyasu:2012fx, Sagayama:2013fna,Asih:2017kp}. 
In recent years, angle-resolved photoemission spectroscopy (ARPES)~\cite{Nakayama:2015jo} and THz optical conductivity~\cite{Sushkov:2015bo} have been used to elucidate the nature of the low-$T$ phase.
As we will see later, however, the Weyl semimetal phase is absent in the LDA+dynamical mean-field (DMFT) phase diagram.

The pyrochlore iridates shown in Fig.~\ref{fig:iridate-pd-exp} contain Ir$^{4+}$ with a 5$d^5$ configuration.
In the single-particle picture, this leaves one hole in the $\jeff$=1/2 band as discussed in section 3.1.
Thus, the electronic properties of the compounds are often discussed based on an effective single-band picture of the $\jeff$=1/2 orbital.
However, to what extent is this simplified description justified in a quantitative study of the properties of the iridates? 
We discuss this point in this review based on the results of first-principles calculations performed by some of the authors and their co-workers~\cite{Shinaoka:2015vma}. 
Our particular focus will be on Y$_2$Ir$_2$O$_7$, which is located in the insulating regime.
Y$^{3+}$ has no $f$ moment, which may make Y$_2$Ir$_2$O$_7$ a prototype compound for studying strong correlations among $5d$ electrons. 
In section~\ref{sec:iridate}, we discuss the experimental phase diagram based on a theoretical phase diagram obtained by LDA+dynamical mean field theory (LDA+DMFT) with respect to the onsite Coulomb interaction $U$ and temperature $T$~\cite{Shinaoka:2015vma}.
\begin{figure}
	\includegraphics[width=.5\textwidth,clip]{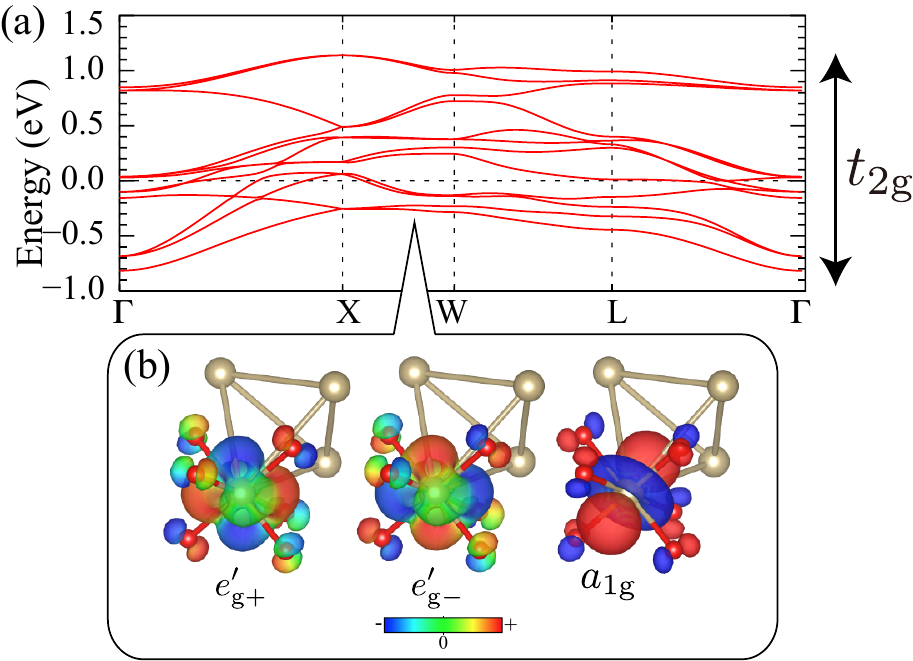}
	\caption{
(a) LDA band structure computed for Y$_2$Mo$_2$O$_7$ and (b) maximally localized Wannier functions constructed from the band structure.
There are twelve $t_\mathrm{2g}$ bands in (a) (each of them is doubly degenerate due to time-reversal symmetry and inversion symmetry).}
	\label{fig:Mo-band}
\end{figure}

\section{Spin-orbital interplay in frustrated Mo pyrochlores}
Figure~\ref{fig:Mo-band}(a) shows the LDA band structure of Y$_2$Mo$_2$O$_7$ in the nonmagnetic state.
The effects of the SOC were included by solving the relativistic Kohn-Sham equation (see section~\ref{sec:method}). 
Isolated states having $t_{2g}$ character are observed in the vicinity of the Fermi level.
The nonmagnetic metallic state is stable in the LDA calculation, which is inconsistent with the insulating nature of the compound.
This can be ascribed to the LDA, which underestimates electron-electron interaction effects. 
Solovyev studied the effect of the intra-atomic Coulomb repulsion (Hubbard $U$) on the Mo sites using the LDA+$U$ method, and showed that the effect of $U$ makes the system insulating~\cite{Solovyev:2003iv}.
A Mo$^{4+}$ ion has an electron configuration of 4$t_\mathrm{2g}^{2}$.
One electron occupies the $a_{1g}$ state which is  half-filled, and the other occupies the $e'_{g}$ states [Fig.~\ref{fig:CF}(a)].
As the $e'_{g}$ states are quarter-filled, one may expect that orbital degrees of freedom are active.
However, the orbital magnetic moment of the $e'_{g}$ states couples to the spin moment through the SOC $\zeta$,
and therefore they do not behave independently.
In what follows, we will see that the orbital degrees of freedom of Mo$^{4+}$ and the spin-orbital physics arising from them play a crucial role in the magnetism of this compound. 

\begin{figure}
	\includegraphics[width=.5\textwidth,clip]{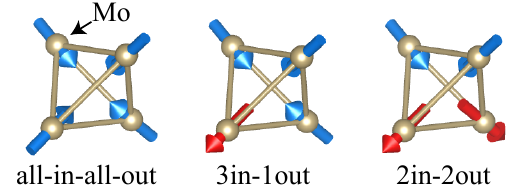}
	\caption{Three competing magnetic structures in Y$_2$Mo$_2$O$_7$.
	}
	\label{fig:Mo-magnetic-structure}
\end{figure}

Some of the authors performed LDA+$U$ calculations with $U$ $\geq$ 3 eV and compared the total energy of the insulating states with different periodic magnetic structures~\cite{PhysRevB.88.174422}.
It was found that three magnetic structures, shown in Fig.~\ref{fig:Mo-magnetic-structure}, are nearly degenerate in energy. 
In these structures, the magnetic moments are localized mostly on the Mo sites, while the O sites contribute only a few percent to the spin moments.
The three states are different, as the sum of the four spins at the corners of a tetrahedron is different: The all-in-all-out structure is an antiferromagnetic state where the four spin moments cancel each other, whereas the spin moments do not cancel in the other two configurations.
Therefore, ferromagnetic and antiferromagnetic correlations are competing in the present system. A detailed analysis using an effective spin model and a multi-orbital Hubbard model revealed that there is keen competition between the 2in--2out and all--in--all--out magnetically ordered states.
We refer the interested readers to Ref.~\cite{PhysRevB.88.174422} for more details.
We also note that another independent DFT+$U$ study revealed a substantial coupling between spin and orbital  degrees of freedom~\cite{Silverstein:2014bka}.
It was also shown that each of these orbital states is accompanied by local lattice distortions, indicating a coupling between lattice and orbital degrees of freedom.

\begin{figure}
	\includegraphics[width=.5\textwidth,clip]{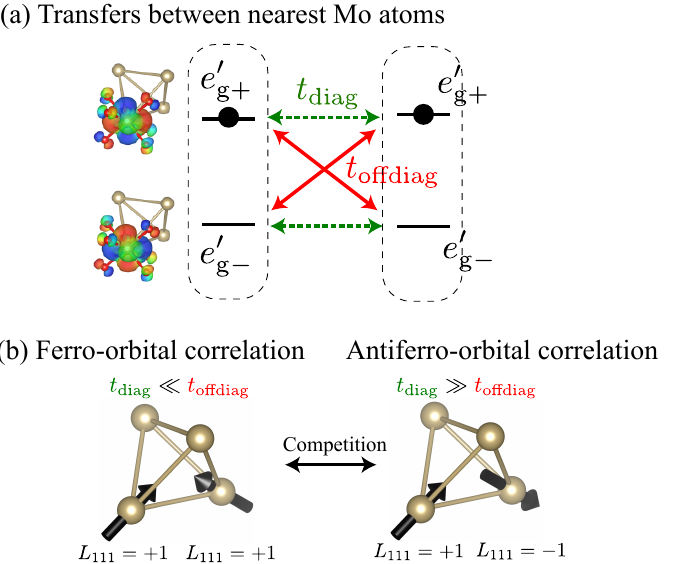}
	\caption{Competing orbital orderings in Y$_2$Mo$_2$O$_7$.
		(a) Hopping between $\egp$ orbitals.
		(b) Two different patterns of orbital ordering are stabilized by changing the ratio $t_\mathrm{offdiag}$/$t_\mathrm{diag}$.
	}
	\label{fig:Mo-orbital-ordering}
\end{figure}

What is the origin of this magnetic competition? As a matter of fact, it is deeply connected to the orbital degrees of freedom which have not been considered thus far. This is schematically explained in Fig.~\ref{fig:Mo-orbital-ordering}. Let us consider the hopping of $e'_{g}$ electrons between two neighboring Mo atoms. A ferro-orbital alignment of the two electrons leads to a gain of kinetic energy when the orbital off-diagonal transfer integral $t_{\rm offdiag}$ dominantes the orbital-diagonal transfer integral $t_{\rm diag}$. In contrast, the antiferro-orbital alignment is favored in the opposite limit, namely when $|t_{\rm offdiag}| \ll |t_{\rm diag}|$ holds. In the case of ferro-orbital alignment, all the orbital magnetic moments on a tetrahedron are inward- or outward-oriented. Since the spin and orbital moments are antiparallel coupled by the SOC, the ferro-orbital alignment favors all-in-all-out magnetic ordering, whereas the antiferro-orbital alignment favors 2in-2out magnetic ordering. 

In order to see whether $t_{\rm offdiag}$ or $t_{\rm diag}$ is dominant in the real material, some of the authors evaluated the transfer integrals between the maximally localized Wannier functions constructed from the LDA solutions [Fig.~\ref{fig:Mo-band}(b)]. It was found that the two terms are close in magnitude. This indicates that ferro- and antiferro-orbital alignments are competing in the present system, which results in a competition between spin structures through the SOC. 

The nature of the spin-glass phase and the role of randomness still remain to be clarified.
Recently, Thygesen and co-workers performed a detailed neutron and x-ray pair-distribution function analysis of the crystal structure of Y$_2$Mo$_2$O$_7$~\cite{2017PhRvL.118f7201T}.
They found that the Mo$^{4+}$ ions displace according to a local 2in--2out rule on each Mo tetrahedron.
This indicates the important role of spin-lattice coupling and that local 2in--2out magnetic structures survive in the spin-glass phase (see Fig.~\ref{fig:Mo-orbital-ordering}).
Long-range ordering may be prevented by the microscopic degeneracy of coverings on the pyrochlore lattice with 2in--2out structures (spin-ice-like degeneracy).
Further theoretical and experimental studies are needed to settle this issue.

\section{Non-collinear magnetism in Os pyrochlore}\label{sec:osmate}
\begin{figure}
	\includegraphics[width=.5\textwidth,clip]{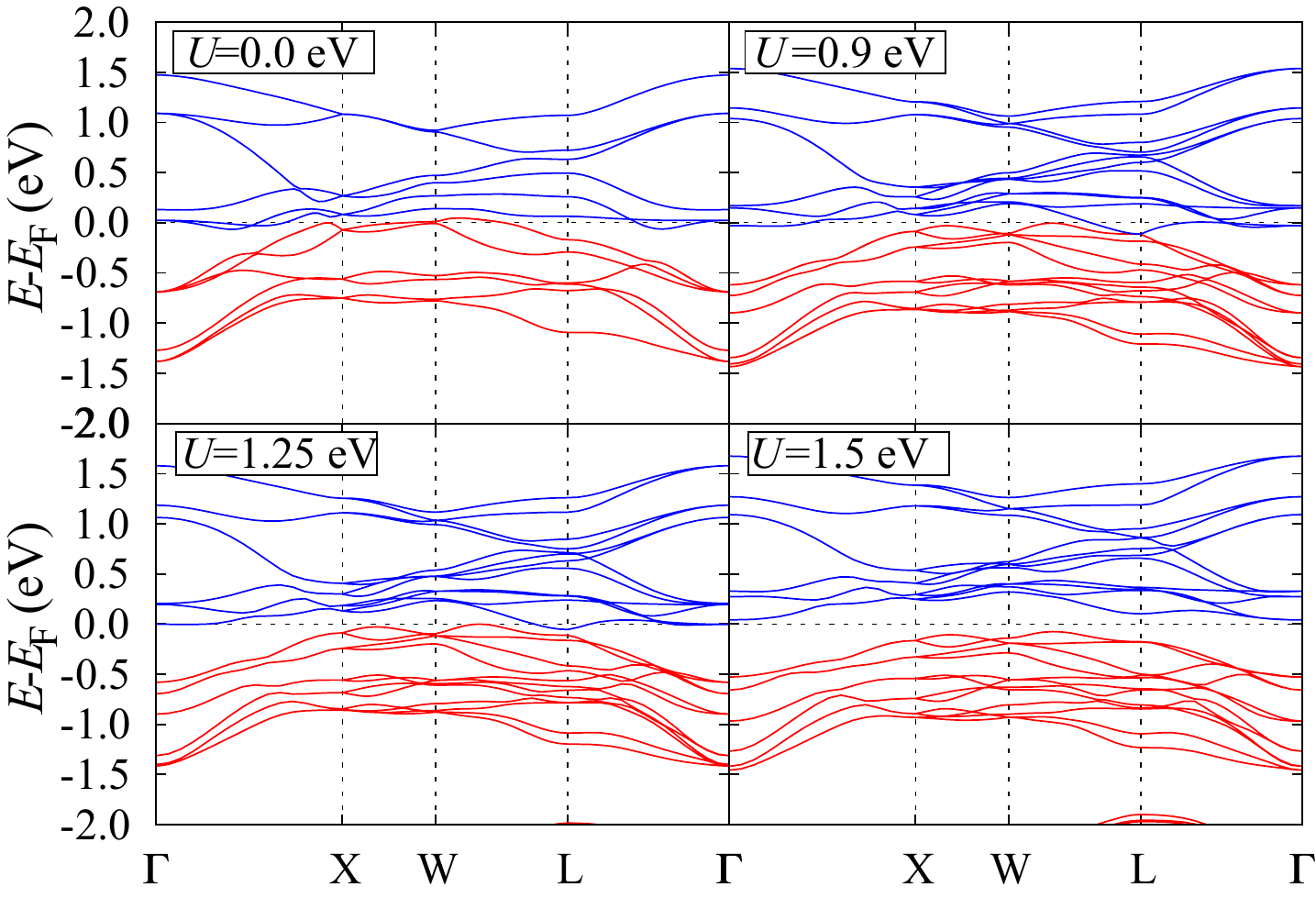}
	\caption{Band structure of Cd$_2$Os$_2$O$_7$ computed by the LDA+$U$ method.
		The result for $U=0$ corresponds to the LDA band structure.
		For $U \ge 0.9$ eV, the ground states are all--in--all--out ordered states.
	}
	\label{fig:Os-band}
\end{figure}

In Fig.~\ref{fig:Os-band}, the band structures of Cd$_2$Os$_2$O$_7$ computed by the LDA$+U$ method are shown. The present result for $U=0$ (corresponding to the LDA) is in good agreement with the previous result reported by Harima~\cite{Harima:2002km}.
Since a unit cell contains four Os atoms, there are 24 $t_{\rm{2g}}$ bands (taking the spin degeneracy into account).
Contrary to the Ir oxide case, to be described below, the splitting into $j_{\rm{eff}} = 1/2$ and $3/2$ bands is not clearly observed.
In this 5$d^3$ situation, the $t_{\rm{2g}}$ bands are half occupied and semi-metal-like features are observed. When applying ``$+U$" similarly to the Mo oxide case, the all-in-all-out magnetic order becomes stable as the ground state for $U \geq 0.9$ eV.
It should be noted that no Brillouin-zone folding occurs since the all-in-all-out magnetic order has a wave vector $\mbox{\boldmath $Q$} = 0$.
As the magnetic moment grows for larger $U$, a small indirect gap opens between the valence and conduction bands as a consequence of the band shift induced by the magnetic ordering.
At each {\boldmath $k$}, however, a larger direct gap exists.
These features are consistent with the distinct gap observed by optical conductivity measurements~\cite{Padilla:2002ep} and the small gap deduced from the temperature dependence of the  resistivity~\cite{Sleight1974,Mandrus:2001ia}.

In the real material, a metal-to-insulator transition occurs as $T$ is lowered.
Although the LDA+$U$ calculations were carried out at $T=0$,
they might give some insight into the nature of the finite-$T$ transition.
The finite-$T$ transition may be caused by shifts of the bands when the magnetic moment increases continuously with decreasing temperature below $T_{\rm{c}}$. Such a transition is called ``Lifshitz transition''.\footnote{Strictly speaking, a Lifshitz transition can be defined only at $T = 0$, where the Fermi surfaces are well defined.}
A recent optical spectroscopy experiment supports this scenario~\cite{Sohn:2015kc}.

In Cd$_2$Os$_2$O$_7$,
the all--in--all--out magnetic order is stabilized by the strong magnetic anisotropy induced by the SOC~\cite{Shinaoka:2012ja}.
Table~\ref{table:os-magnetic-anisotropy} shows the values of the nearest-neighbor exchange coupling ($J$), the single-ion anisotropy ($\Asia$)
and the Dzyaloshinskii-Moriya (DM) interaction ($\ADM$) estimated by LDA+$U$ calculations.
The exact form of the effective spin model is given by Eq.~(1) in Ref.~\cite{Shinaoka:2012ja}.
There exists a 
strong magnetic anisotropy in this compound even near the metal-insulator transition: $\Asia>0$ corresponds to an easy-axis anisotropy, while $\ADM > 0$ stabilizes the all--in--all--out order in combination with the antiferromagnetic $J~(>0)$.

Our LDA$+U$ studies were done around 2012. In those days, it was technically difficult to perform finite-temperature calculations. As described in the next section, finite-temperature calculations including electron correlation effects have in the meantime become possible even for 5$d$ transition-metal oxides using the LDA+DMFT method. In parallel with the improvement of experimental techniques such as angle-resolved photoemission spectroscopy (ARPES)~\cite{Nakayama:2015jo}, modern first-principles calculations initiate a new era in the study of finite-temperature phase transitions. Do quasi-particle bands survive close to $T_{\rm{c}}$? What is the effect of thermal fluctuations of the magnetic moment? It is expected that these and related questions will be solved in the near future. Understanding
the bulk properties of 
Cd$_2$Os$_2$O$_7$
%this compound 
is still a hot research subject~\cite{Hiroi2015,Sohn:2015kc,Vale:2016bw,Nguyen:2017hf}.

\if0
\begin{figure}
	\centering
	\includegraphics[width=.475\textwidth,clip]{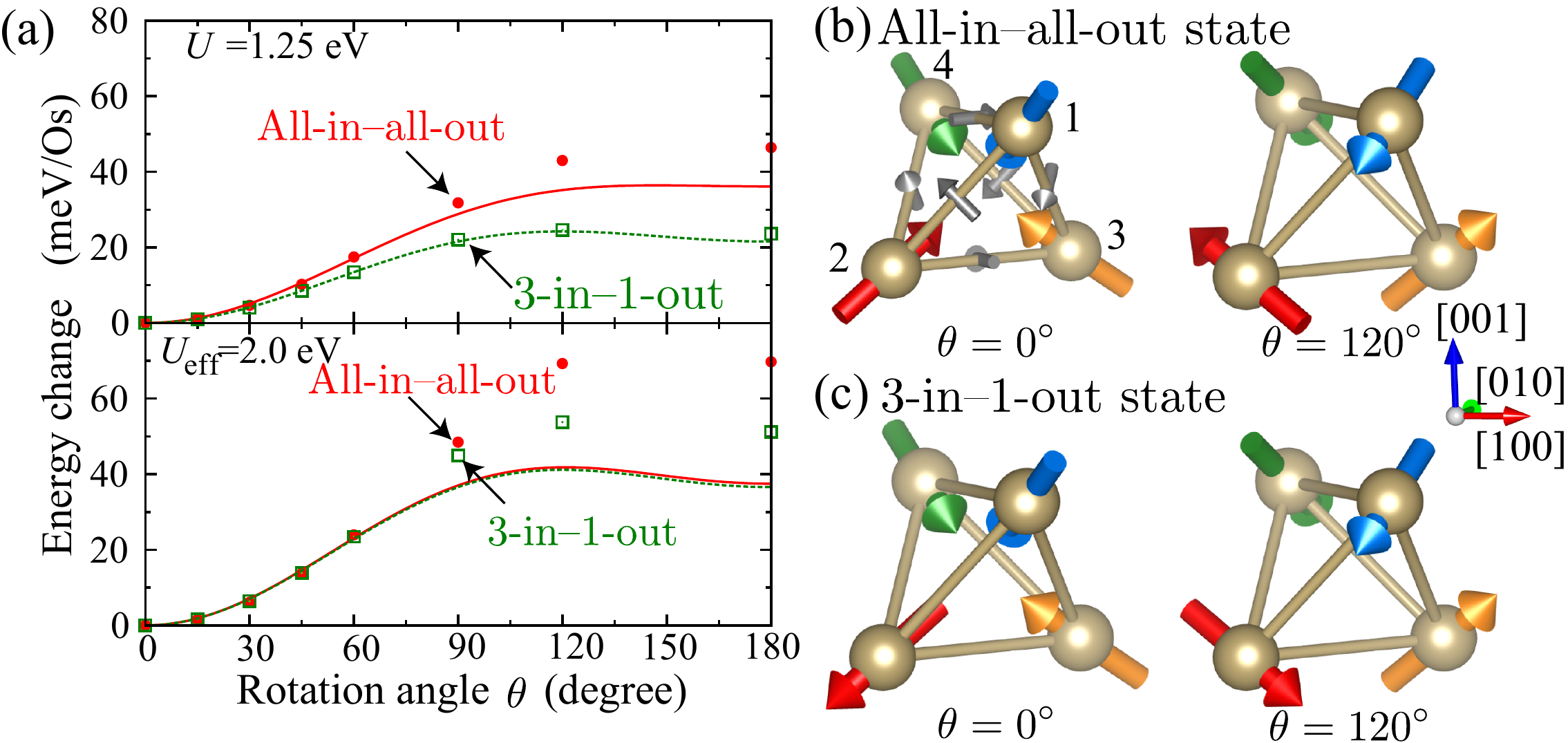}
	\caption{(color online).
		(a) Magnetic anisotropy energies estimated by the rotation of Os magnetic moments around the [001] axis.
		(b)/(c) all-in/all-out and 3-in/1-out ordered states rotated by $\theta=0^\circ$ and $\theta=120^\circ$ around the [001] axis.
		Every spin is perpendicular to the local $\langle 111 \rangle$ axis at $\theta=120^\circ$.
		In (b), the small (gray) vectors and the labels represent the direction vector $\vec{d}_{ij}$ of the DM interaction and the site indices $i$, respectively.
	}
	\label{fig:os-magnetic-anisotropy}
\end{figure}
\fi
\begin{table}
	\centering
	\begin{tabular}{lccc}
		\hline
		$\Ueff$ (eV) & $J$ (meV) & $\Asia$ (meV) & $\ADM$ (meV) \\
		\hline
		1.25 &   14 &  24 & 4 \\
		2.0  &   35 &  41 & 0 \\
		\hline
	\end{tabular}
	\caption{
		Nearest-neighbor exchange coupling, single-ion anisotropy and Dzyaloshinskii-Moriya (DM) interaction estimated by LDA+$U$ calculations
		for Cd$_2$Os$_2$O$_7$ and indicated values of $\Ueff \equiv \bar{U}-\bar{J}$ [see Eq.~(\ref{eq:dudarev})].
		The direction vectors of the DM interaction are shown in Fig.~4(b) of Ref.~\cite{Shinaoka:2012ja}.
	}
	\label{table:os-magnetic-anisotropy}
\end{table}

\section{Mott physics and magnetism in Ir pyrochlores}\label{sec:iridate}
\subsection{Phase diagram at half filling}\label{sec:iridate-pd}
The crossover between a paramagnetic metal and a paramagnetic insulator in Fig.~\ref{fig:iridate-pd-exp} clearly indicates the importance of Mott physics in understanding the electronic properties of pyrochlore iridates.
Mott physics in real compounds can be (partially) described by the LDA+DMFT method.
As reviewed in section~\ref{sec:method}, the LDA+DMFT method can capture local correlation effects,
%where the dynamical nature of *** plays a substantial role.
including dynamical fluctuations. 
The LDA+DMFT method has been used to study various transition metal oxides (for a review, refer to Ref.~\cite{Kotliar:2006fl}).
However, the applicability of the method is still limited to a small class of compounds.
When studying finite-$T$ properties, the LDA+DMFT method is typically used in combination with continuous-time QMC calculations \cite{Gull:2011jda}. 
However, these CT-QMC calculations suffer from a sign problem at low temperature when multi-orbital aspects of the compound become relevant.
In particular, the computation of magnetism in compounds with strong SOC still remains a challenging problem.

\begin{figure}
	\includegraphics[width=.5\textwidth,clip]{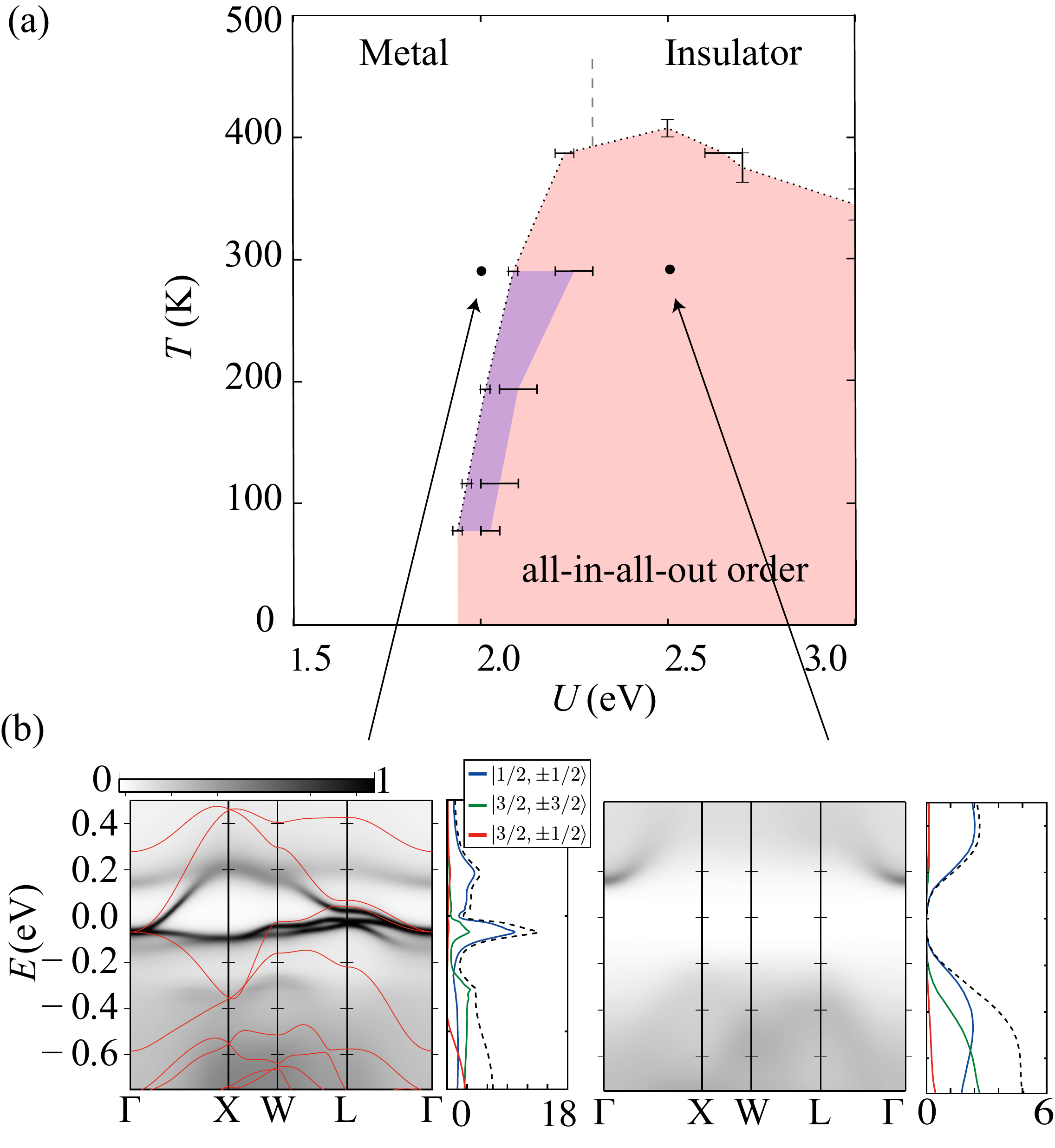}
	\caption{Phase digram of Y$_2$Ir$_2$O$_7$ computed by LDA+DMFT [(a)]. The real compound may be located at $U\simeq 2.5$~eV. (b) Momentum-resolved spectral function computed at 290 K. The spectral functions projected onto the $\jeff$ basis ($|\jeff, \jeff^{111}\rangle$) are also shown.}
	\label{fig:iridate-pd}
\end{figure}

\begin{figure}
	\includegraphics[width=.4\textwidth,clip]{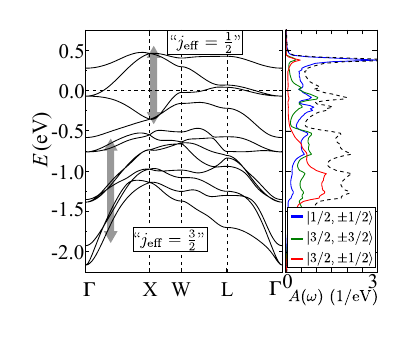}
	\caption{LDA band structure computed for Y$_2$Ir$_2$O$_7$. The density of states projected onto the $\jeff$ basis is also shown.}
	\label{fig:iridate-lda}
\end{figure}

Figure~\ref{fig:iridate-pd} shows a theoretical phase diagram computed by DMFT calculations based on the LDA band structure of Y$_2$Ir$_2$O$_7$ (see Fig.~\ref{fig:iridate-lda})~\cite{Shinaoka:2015vma}.
Some of us systematically explored the phase diagram by changing $U$ (the Hund coupling parameter was fixed at $J_\mathrm{H}/U=1/10$).
The large-$U$ region in Fig.~\ref{fig:iridate-pd} corresponds to the small $A$-site ionic radius region 
in the experimental phase diagram~(Fig.~\ref{fig:iridate-pd-exp}).
There is a dome-shaped all-in–all-out magnetically ordered phase at large $U$,
which extends up to about 400 K.
It is empirically known that DMFT calculations for 3D compounds overestimate magnetic transition temperatures typically by a factor of 2--3 because of the neglect of spatial fluctuations.
Considering this point, the experimental and theoretical phase diagram show a good agreement.


The broken line above the magnetic transition temperature indicates the crossover between a paramagnetic metal and a paramagnetic insulator.
This can be identified by the gradual decrease of the spectral weight near the Fermi level as one moves from the small-$U$ region to the large-$U$ region. 
In the paramagentic insulating region, the DC resistivity is expected to show an insulating behavior above the magnetic transition temperature.
The target compound Y$_2$Ir$_2$O$_7$ exhibits insulating behavior of the resistivity at room temperature.
Thus, $U \simeq 2.5$ eV may be a realistic value for the compound.

The lower panel of Fig.~\ref{fig:iridate-pd} shows the momentum-resolved spectral function computed at 290~K.\footnote{To compute the spectral function, the (local) DMFT self-energy was continued to the real-frequency axis and then inserted into the Dyson equation, whose solution yields the lattice Green's function on the real-frequency axis.}.
In the paramagnetic metallic region, quasi-particle bands of the $\jeff$=1/2 manifold are clearly observed.
This $\jeff$=1/2 manifold shows a substantial band narrowing from the LDA value of $\simeq$ 1 eV down to approximately 0.4 eV. 
In the insulating regime, the quasi-particle bands of the $\jeff$=1/2 manifold are smeared out, indicating its Mott nature.

Here comes an important question.
How \textit{relevant} is the role of the $\jeff$=3/2 manifold for $E<0$?
To answer this question, we first focus on the energy region ($E<-0.6$ eV) where the $\jeff$=3/2 bands are originally located before $U$ is turned on.
Even in the paramagnetic metallic region, these bands are smeared out and cannot be clearly identified.
Another interesting observation is that the low-energy states consist mainly of the $\jeff$=1/2 orbitals for $\omega > 0$, while in LDA the $\jeff$=3/2 manifold contributes significantly to the density of states near $\omega=0$ (see Fig.~\ref{fig:iridate-lda}).
A similar purification of the spectral function was found in the LDA+DMFT study of Sr$_2$IrO$_4$~\cite{Zhang:2013iy}.

These results indicate that the effects of the self-energy (refer to section~\ref{sec:method}) on the $\jeff$=3/2 manifold is substantial,
which has nontrivial consequences. 
In LDA+DMFT calculations for the pyrochlore iridates,
one can construct an effective one-band model for the $\jeff$=1/2 manifold,
where the $\jeff$=3/2 manifold is neglected and is treated as frozen states.
We confirmed that the dome-shaped all--in--all-out region is reduced in height by about a factor of two~\cite{singlesite}.
This indicates that the multi-orbital aspects may stabilize the all--in--all--out magnetic ordering, and thus, the three-orbital description is necessary for a quantitative understanding of this compound.

As mentioned in Sec.~3.4,
previous LDA+$U$ studies indicated the existence of a Weyl semimetallic phase between a magnetically ordered insulating phase and a paramagnetic metallic phase~\cite{Wan:2011hi,Ishii:2015kw}.
The width of the Weyl semimetallic phase was estimated to be around 0.2 eV~\cite{Ishii:2015kw}.
In contrast, in the LDA+DMFT phase diagram,
no intermediate phase was found between the all--in--all--out magnetically ordered phase and the paramagnetic phase, which are separated by a first-order transition line at low $T$.
This may indicate that the Weyl semimetallic phase is less stable under the influence of strong correlations (e.g., strong band-width renormalization).
However, we must keep in mind that non-local correlations are neglected in single-site DMFT calculations, which may have a substantial effect near the magnetic quantum critical point.
Therefore, the existence of such a non-trival phase near the magnetic critical point still remains an open issue.

\begin{figure}[b]
	\centering\includegraphics[width=.4\textwidth,clip]{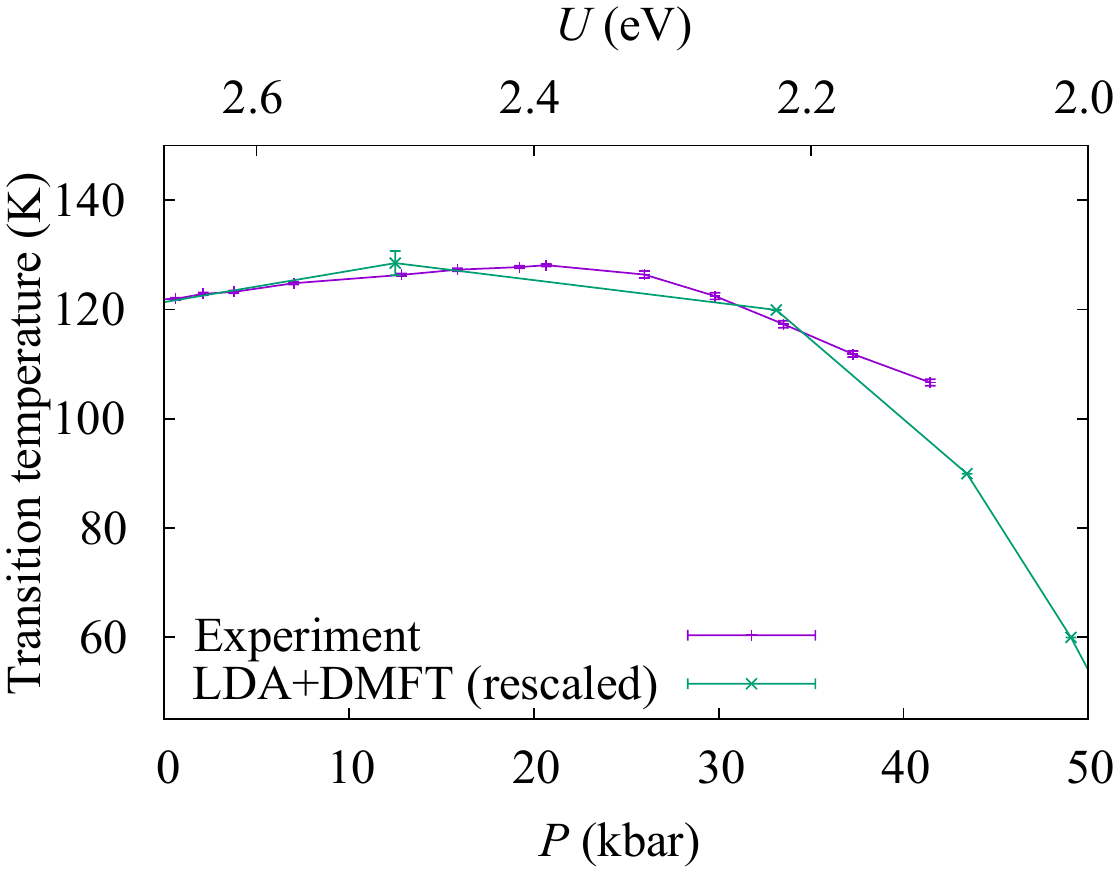}
	\caption{
		(Color online) Comparison of the pressure dependence of the critical transition temperature of Eu$_2$Ir$_2$O$_7$ and the LDA+DMFT results.
		The transition temperatures of LDA+DMFT were scaled by a factor of 0.31.
		We extracted the experimental data from Fig.~6 in Ref.~\cite{Prando:2016cn}.
	}
	\label{fig:Eu}
\end{figure}

Recently, Prando \textit{et al}. reported the magnetic properties of a related compound Eu$_2$Ir$_2$O$_7$ 
under hydrostatic pressure by macroscopic and local-probe techniques~\cite{Prando:2016cn}.
They found that the magnetic transition temperature increases up to $P$ = 20 kbar and then drops as hydrostatic pressure is further increased.
They compared their results with our theoretical results under the assumption that the hydrostatic pressure corresponds to changing the $U/W$ ratio ($W$ is the band width).
Figure~\ref{fig:Eu} shows a similar plot~\cite{Shinaoka:2017ez}, where the theoretical transition temperature is scaled by a factor of 0.31 to correct the overestimation by the local approximation.
One can see that the $P$ dependence of the data is consistent with the theoretical curve, although experiments for higher $P$ are needed for a more detailed comparison.
While this comparison assumes that $U/W$ changes as $P$ is increased,
it remains to be clarified which microscopic parameters actually vary (see the related discussion in section~\ref{sec:summary}).

\subsection{Doped compounds}
We next discuss the effects of carriers in pyroclore iridates.
An interesting observation in the spectral function computed by the DMFT method is the existence of nearly flat bands slightly above and below the Fermi level [see Fig.~\ref{fig:iridate-pd}(b)].
These may get close to the Fermi level as the compound is doped with holes or electrons.
Such effects were investigated by performing LDA+DMFT calculations with different numbers of electrons (without changing the non-interacting band structure)~\cite{Shinaoka:2015vma}.
The LDA+DMFT calculations indicated the emergence of non-Fermi-liquid-like behavior down to low $T$ (the lowest $T$ computed was around 50~K) in the hole-doped cases.
This can be understood in terms of long-lived local magnetic moments induced by the nearly flat bands down to low $T$.

Recently, hole-doped derivatives of Eu$_2$Ir$_2$O$_7$ were investigated through transport and magnetic measurements~\cite{Banerjee:2017de}.
The metal-insulator transition temperature was found to get substantially reduced with hole doping,
and for 10\% Sr doping the composition is metallic down to temperatures as low as 5 K.
This study found that these doped compositions violate the Mott-Ioffe-Regel condition for minimum electrical conductivity and show a distinct signature of non-Fermi liquid behavior at low temperature.
The authors attributed this peculiar behavior to a disorder-induced variation
of the spin-orbit-coupling parameter between Ir$^4+$ and Ir$^5+$ ions.
However, the nearly flat bands may provide an alternative explanation.

\section{Theoretical framework: LDA+$U$ and LDA+DMFT}\label{sec:method}
In the previous sections, we have discussed rich phenomena arising from spin-orbital physics in $4d$ and $5d$ pyrochlores.
The description of these phenomena such as non-collinear magnetism and Mott transitions requires the use of formalisms which go beyond the  LDA.
In this section, we provide a brief description of these methods for a comprehensive understanding of the technical advances underlying our studies.
First, in section~\ref{sec:LDAU}, we provide a brief description of the LDA+$U$ method, which was used to investigate the molybdates and osmate.
In section~\ref{sec:wannier}, we explain the theory of maximally localized Wannier functions, on which all the studies shown in this review are based. 
In addition to these general theories, we also explain a convenient local coordinate system for pyrochlore oxides.
These two sections furthermore serve as an appetizer for the detailed discussion of the LDA+DMFT method in section~\ref{sec:LDADMFT}.
In section~\ref{sec:LDADMFT}, we describe the basic theory of LDA+DMFT and some technical details especially on solving multi-orbital quantum impurity models.
The solution of the DMFT impurity problem is one of the biggest bottlenecks in LDA+DMFT calculations for $4d$ and $5d$ systems due to a severe negative sign problem resulting from the SOC in Monte Carlo based techniques.
We describe some recipe for avoiding a severe sign problem. 

\subsection{LDA+$U$}\label{sec:LDAU}
Despite its success in the quantitative description of a wide range of materials, DFT with common approximations for the exchange-correlation functional, such as the local density approximation (LDA) and generalized gradient approximation (GGA), cannot properly capture electron-electron interaction effects in strongly correlated materials. A typical example is La$_2$CuO$_4$, a mother compound of a copper-oxide superconductor. It is experimentally an antiferromagnetic insulator, whereas DFT-LDA yields a nonmagnetic metallic solution. Similar qualitatively wrong results are often obtained in transition-metal oxides. The error can be ascribed to the approximate exchange-correlation functional, and also to the local and static effective potential in the Kohn-Sham framework. This drawback is cured by explicitly introducing an on-site Coulomb repulsion (Hubbard $U$) \cite{anisimov1991band,anisimov1997first}. Let $\{ n_{mm'}^{\sigma} \}$ be the density matrix of the interacting localized orbitals, e.g. the 3$d$ orbitals. In the LDA+$U$ method, the total energy functional is written as

\begin{equation}
E^{{\rm LDA}+U} = E^{\rm LSDA} + E^{U}[\{  n_{mm'}^{\sigma}\}] - E^{\rm dc}[\{  n_{mm'}^{\sigma}\}] \;,
\end{equation}
where $E^{\rm LSDA}$ is the total energy functional in the local spin-density approximation (LSDA),
$E^{U}$ represents the contribution from the Hubbard-type interaction, and $E^{\rm dc}$ is the double-counting term.
Since the effects of $E^{U}$ are partly included in $E^{\rm LSDA}$, we have to remove the double counting by subtracting $E^{\rm dc}$. There are several proposals for the explicit form of $E^{U}$ and $E^{\rm dc}$. One is a rotationally-invariant form by Liechtenstein {\it et al.} \cite{liechtenstein1995density}, 
\begin{eqnarray}
E^{U} &=&  \frac{1}{2} \sum_{\{m\},\sigma} \{ \langle m, m'' | V_{ee} | m', m''' \rangle 
n_{mm'}^{\sigma} n_{m''m'''}^{-\sigma} \nonumber \\
&& + ( \langle m, m'' | V_{ee} | m', m''' \rangle \nonumber \\
&& - \langle m, m'' | V_{ee} | m''', m' \rangle ) 
n_{mm'}^{\sigma} n_{m''m'''}^{\sigma}  \}
\; ,\\
E^{\rm dc} &=& \frac{1}{2} U n (n-1) 
- \frac{1}{2}J 
[ n^{\uparrow} (n^{\uparrow}-1)
n^{\downarrow} (n^{\downarrow}-1) ] \;,
\end{eqnarray}
where $n^{\sigma} = \Tr(n_{m m'}^{\sigma})$, $n=n^{\uparrow}+n^{\downarrow}$, and 
$U$ and $J$ are screened Coulomb and exchange interaction parameters, respectively.
The corresponding effective one-electron potential becomes orbital-dependent:

\begin{eqnarray}
V_{mm'}^{\sigma} &=&  \sum_{m'',m'''} \{ \langle m, m'' | V_{ee} | m', m''' \rangle 
n_{m''m'''}^{-\sigma} \nonumber \\
&& + ( \langle m, m'' | V_{ee} | m', m''' \rangle \nonumber \\
&& - \langle m, m'' | V_{ee} | m''', m' \rangle ) 
n_{m''m'''}^{\sigma}  \} \nonumber \\
&& -U \Big(n-\frac{1}{2}\Big) + J\Big(n^{\sigma}-\frac{1}{2}\Big) \;.
\end{eqnarray}
Solovyev \textit{et al.}~\cite{Solovyev94} proposed a simpler form where the Coulomb interaction is spherically averaged.
In our first-principles studies of the Os and Mo pyrochlores~\cite{PhysRevB.88.174422,Shinaoka:2012ja},
we used 
\begin{eqnarray}
& E^{{\rm LDA}+U} =E^{\rm LSDA} +\nonumber\\
&\hspace{1em}  \frac{\bar{U}-\bar{J}}{2}\sum_{\sigma} 
\Big[ \sum_{m}n_{mm}^{\sigma} - \sum_{mm'}n_{mm'}^{\sigma} n_{m'm}^{\sigma} \Big] \;,\label{eq:dudarev}
\end{eqnarray}
where $\bar{U}$ and $\bar{J}$ are spherically-averaged screened Coulomb and exchange interaction parameters, respectively~\cite{dudarev1998electron}.

Note that the relativistic effect including the SOC can be fully considered in solving the relativistic Kohn-Sham equation~\cite{Oda98,Kosugi:2011gma}.  
The LDA+$U$ method can be regarded as a hybrid method in the sense that the DFT-LDA total energy is supplemented by the many-body terms associated with $U$ and $J$ in the Hartree-Fock approximation. The results depend on the choice of $U$ and $J$.
Although they are simply treated as adjustable parameters in many cases, there exist a few theoretical frameworks for evaluating them from first-principles, such as the constrained LDA \cite{gunnarsson1989density,gunnarsson1990gunnarsson}, linear-response approach~\cite{cococcioni2005linear}, and constrained RPA method \cite{aryasetiawan2004frequency}. 

\subsection{Construction of Wannier functions and local coordinate system}\label{sec:wannier}
Localized orbitals need to be constructed for the formulation of low-energy effective tight-binding models and the implementation of methods like LDA+DMFT which are based on such effective tight-binding models.
First-principles schemes based on localized orbitals as basis functions (LMTO and FLAPW \textit{etc}.) %, these orbitals can be directly 
directly provide the orbitals 
used in the LDA+DMFT calculations.
A more general framework, which is applicable also to planewave methods, is the construction of Wannier functions. The Wannier function $\varphi_{n\mathbf{R}} ({\bf r})$ associated with a Bloch function  $\psi_{n\mathbf{k}}({\bf r})$ is defined as  

\begin{equation}
\varphi_{n\mathbf{R}} ({\bf r}) = \frac{V}{(2\pi)^{3}} \int e^{-i\mathbf{k}%
\cdot\mathbf{R}} \psi_{n\mathbf{k}}({\bf r}) d^{3}k\;,
\label{eq:wannier}%
\end{equation}
where $n$ is the orbital index, ${\bf R}$ is the cell index, and ${\bf k}$ is the wave vector. This Wannier function is not uniquely defined because the Bloch function has gauge degrees of freedom, $\psi_{n\mathbf{k}}({\bf r}) \rightarrow e^{i \theta({\mathbf k})}\psi_{n\mathbf{k}}({\bf r})$. We may thus generalize the definition of the Wannier function as

\begin{eqnarray}
\varphi_{n\mathbf{R}} ({\bf r}) &=&  \frac{V}{(2\pi)^{3}} \int e^{-i\mathbf{k}%
\cdot\mathbf{R}} \tilde{\psi}_{n\mathbf{k}}({\bf r}) d^{3}k\;, 
\label{eq:wannier2} \\
\tilde{\psi}_{n\mathbf{k}}({\bf r}) &=& \sum_{m} U_{mn}({\bf k}) %\psi_{n\mathbf{k}}({\bf r}) 
\psi_{m\mathbf{k}}({\bf r})\;.
\label{eq:wannier3}
\end{eqnarray}
Namely, the Bloch functions are reconstructed by taking linear combinations % between them. 
of different bands. 
This is natural, because bands cross each other in ${\bf k}$-space, and hence there is no reason to limit $n$ in Eq.~(\ref{eq:wannier2}) to a single band index throughout the whole Brillouin zone. 

The maximally localized Wannier functions (MLWF) \cite{marzari1997maximally,souza2001maximally,marzari2012maximally} utilize the gauge degrees of freedom, and determine the $U_{mn}({\bf k})$ matrix such that the spread of the Wannier functions, defined by 

\begin{equation}
\Omega = \sum_{n} [ \langle \varphi_{n0} | r^2 | \varphi_{n0} \rangle - 
 \langle \varphi_{n0} | r | \varphi_{n0} \rangle^2 ] \;,
 \label{eq:spread}
\end{equation}
is minimized. Marzari and Vanderbilt have developed the formalism and a practical procedure to minimize the spread by the steepest descent method \cite{marzari1997maximally}. It was extended by Souza \textit{et al.} to the case of  entangled bands,
where the Hilbert space spanned by the MLWF's is determined in such a way that $\tilde{\psi}_{n\mathbf{k}}({\bf r})$ is as smooth as possible in ${\bf k}$-space~\cite{souza2001maximally}. 

The MLWF method has been applied to strongly-correlated systems in Refs.~\cite{miyake2008screened,nakamura2008ab,miyake2009ab,miyake2010comparison}. In these studies, the MLWFs were used to define interacting localized orbitals, and $U$ and $J$ on the MLWFs are computed from first-principles in the constrained RPA~\cite{aryasetiawan2004frequency}.

%Since MLWFs are constructed by minimizing their spread,
In practical calculations, one provides an initial guess (initial orbitals) for the optimization of MLWFs.
This may define the local coordinate system of the resulting tight-binging model.
In our calculations for the Mo and Ir pyrochlores,
for convenience, 
we used $d_{xy}$, $d_{yz}$, $d_{zx}$--like orbitals defined as initial orbitals in the local coordinate system with $z$ aligned along [111], [-111], [1-11], or [11-1] 
on each of four Mo/Ir atoms [see Fig.~\ref{fig:crystal-struc}(b)]. 
They are fully spin-polarized along the local $z$ axis. The resulting tight-binding models are equivalent for each of the four atoms.
To avoid any symmetry breaking in optimizing the Wannier functions,
we projected the initial orbitals to the Bloch wave functions and did not further minimize the spread.
Wannier functions constructed in this way are sometimes called ``one-shot WFs".

In the studies reviewed in this article, we took into account the effects of the SOC by constructing
Wannier functions based on relativistic LDA calculations.
In other words, the SOC appears in the tight-binding models in the form of single-particle terms.

\subsection{LDA+DMFT}\label{sec:LDADMFT}
\begin{figure}[t]
	%\centering
	\includegraphics[width=.5\textwidth,clip]{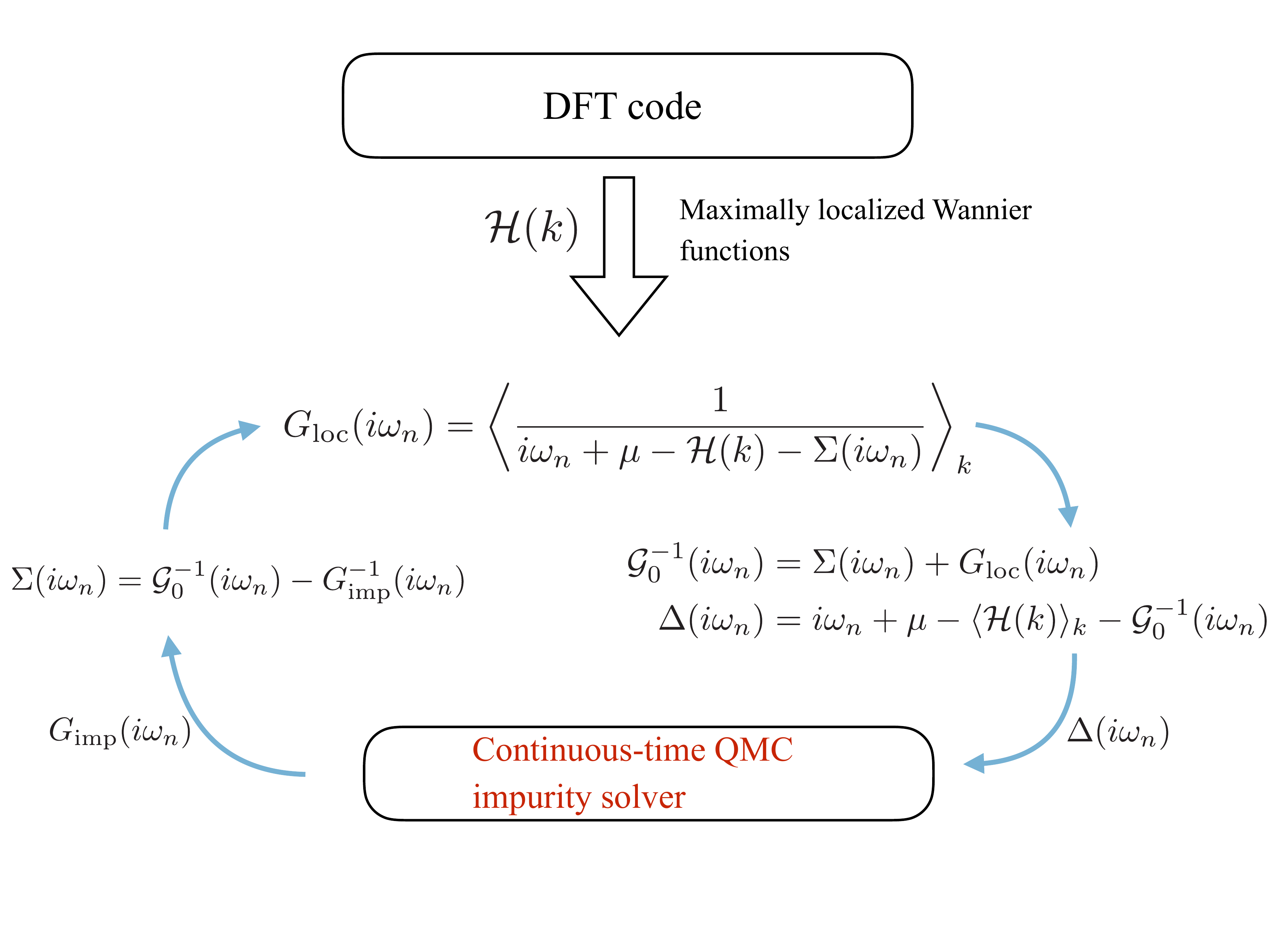}
	\caption{
		(Color online) Illustration of the DMFT self-consistency cycle. The most numerically expensive step is the solution of the impurity problem, i.e., the calculation of $G_{\rm imp}$ for a given $\Delta$. $\braket{\dots}$ denotes an average over momentum space, and $\mu$ is the chemical potential.
	}
	\label{fig:sc_cycle}
\end{figure}

A more accurate treatment of local correlation effects is possible within the LDA+DMFT framework \cite{Held:2001cv,Kotliar:2006fl}, where the lattice system is mapped onto a self-consistently determined quantum impurity model. For this construction, we have to define localized (Wannier) orbitals for the $t_{2g}$ states, and the corresponding local interaction 
\begin{equation}
H_{\rm int}=\frac{1}{2}\sum_{\alpha,\beta,\gamma,\delta,\sigma,\sigma'} U_{\alpha\beta\gamma\delta}d^\dagger_{\alpha\sigma}d^\dagger_{\beta\sigma'}d_{\gamma\sigma'}d_{\delta\sigma},
\end{equation}
for which one typically considers the Slater-Kanamori form $U_{\alpha\alpha\alpha\alpha}=U$, $U_{\alpha\beta\alpha\beta}=U-2J_\mathrm{H}$, $U_{\alpha\beta\beta\alpha}=U_{\alpha\alpha\beta\beta}=J_\mathrm{H}$ ($\alpha\neq \beta$). Here,  
$\alpha$ and $\sigma$ are orbital and spin indices, while $U$ and $J_\mathrm{H}$ denote the on-site repulsion and the Hund coupling, respectively. The self-consistency loop (Fig.~\ref{fig:sc_cycle}) involves the calculation of the impurity Green's function $G_{\rm imp}$ and self-energy $\Sigma$ for a given hybridization function $\Delta$ (or ``Weiss" Green's function $\mathcal{G}_0$), the approximation of the lattice self-energy by the impurity self-energy (DMFT approximation), the calculation of the lattice Green's function for this approximate local self-energy, and finally the identification of the local lattice Green's function $G_{\rm loc}$ with the impurity Green's function $G_{\rm imp}$ (DMFT self-consistency condition). The last step allows to define the new hybridization function $\Delta$ (or $\mathcal{G}_0$) and this loop is iterated until convergence is reached.
If the simulation is restricted to the $t_{2g}$ subspace, %there is no double-counting of correlation energy -- 
and an orbital-independent double counting is adopted,  
it is sufficient to adjust the chemical potential to assure the proper filling.
For more details on the DMFT self-consistent calculation, refer to Refs.~\cite{Georges:1996un,Kotliar:2006fl}.

In the presence of SOC, one needs to treat complex and off-diagonal hybridization functions.
As we will show later, one cannot diagonalize the hybridization functions by a basis transformation with respect to spin and orbital for pyrochlore oxides.
The unbiased numerical solution of such multi-orbital impurity models has only become possible thanks to recent methodological developments.   

\subsubsection{Solution of the impurity problem}
We briefly discuss the hybridization expansion continuous-time Monte Carlo technique, which enables a numerically exact LDA+DMFT simulation of pyrochlore iridates down to a temperature of the order of 100 K. 

Continuous-time impurity solvers~\cite{Gull:2011jda} are based on an expansion of the partition function into a series of diagrams and the stochastic sampling of collections of these diagrams.
The partition function of the impurity model is expressed as a sum (or, more precisely, integral) of the contributions  $w_C$ from all configurations $C$ as
\begin{equation}
Z=\sum_C w_C,
\label{Z_c}
\end{equation}
where $w_C$ can be complex.
Note that the Hermitian property of the Hamiltonian ensures that $Z$ is real after the summation over $C$.
This integral is evaluated by a random walk in configuration space $C_1\rightarrow C_2\rightarrow C_3\rightarrow \ldots$ with weight $|w_C|$.
The random walk needs to be implemented in such a way that each configuration can be reached from any other in a finite number of steps 
({\it ergodicity}) and that {\it detailed balance} is satisfied: 
$|w_{C_1}| p(C_1\rightarrow C_2) = |w_{C_2}| p(C_2\rightarrow C_1)$.
This assures that each configuration $C$ is generated with a probability proportional to $|w_C|$ and yields an estimate for some observable $A$ based on a simple average over a finite number $N$ of measurements:
\begin{eqnarray}
\langle A\rangle &=&\frac{\sum_C w_C A_C}{\sum_C w_C}=\frac{\sum_C |w_C| e^{i\varphi_C} A_C}{\sum_C |w_C|e^{i\varphi_C}}\nonumber\\
&\approx& \frac{\sum_{i=1}^N e^{i\varphi_{C_i}} A_{C_i}}{\sum_{i=1}^N e^{i\varphi_{C_i}}}=\frac{\langle e^{i\varphi} \cdot A\rangle_{\rm MC}}{\langle e^{i\varphi}\rangle_{\rm MC}},
\label{phase}
\end{eqnarray}
where $A_C$ is the value of the observable associated with the configuration $C$, $w_C=|w_C|e^{i\varphi_C}$ is split into the modulus and phase, 
and we assumed that $A_C$ and $w_C$ can be obtained by sampling the same configuration space.  
The error on this Monte Carlo estimate decreases like $1/\sqrt{N}$. 

We focus here on general impurity models with Hamiltonian
\begin{equation}
H=H_\mathrm{loc}+H_\mathrm{bath}+H_\mathrm{hyb},
\label{H}
\end{equation}
where $H_\mathrm{loc}$ describes the impurity, characterized by a small number of degrees of freedom (spin and orbital degrees of freedom denoted collectively by $a$, $b$, $\ldots$), and $H_\mathrm{bath}$ describes an infinite reservoir of free electrons,  labeled by a continuum of quantum numbers $p$ and a discrete set of quantum numbers $\alpha$. Finally, $H_\mathrm{hyb}$ describes the exchange of electrons between the impurity and the bath in terms of hybridization amplitudes $V^{a\alpha}_{p}$.  Explicitly, the three terms are
\begin{eqnarray}
H_\mathrm{loc}&=&\sum_{ab}\epsilon^{ab}_{}d^\dagger_a d^{}_b +\frac{1}{2}\sum_{abcd}U^{abcd}_{}d^\dagger_a d^\dagger_b d^{}_c d^{}_d , \label{Himp} \\
H_\mathrm{bath}&=&\sum_{p\alpha}\varepsilon_{p\alpha}^{} c^\dagger_{p\alpha}c_{p\alpha}^{} , \label{Hbath} \\
H_\mathrm{hyb}&=&\sum_{pa\alpha}\left[V^{a\alpha}_{p} d^\dagger_a c_{p\alpha}^{}+(V^{a\alpha}_{p})^* c_{p\alpha}^{\dagger}d^{}_a \right].\label{Hmix}
\end{eqnarray}
The hybridization-expansion approach \cite{Werner:2006ko,Werner:2006iz} is based on an expansion of the partition function $Z$ in powers of the impurity-bath hybridization term,  $H_{\rm hyb}$, and an interaction representation in which the time evolution is determined by $H_{\rm loc}+H_{\rm bath}$: $O(\tau)=e^{\tau(H_{\rm loc}+H_{\rm bath})}Oe^{-\tau(H_{\rm loc}+H_{\rm bath})}$.
  
Since $H_\mathrm{hyb} \equiv H_\mathrm{hyb}^{d^\dagger}+H_\mathrm{hyb}^d=\sum_{pa\alpha} V_p^{a\alpha} d^\dagger_a c^{}_{p\alpha} + \sum_{pa\alpha} (V_p^{a\alpha})^* c^\dagger_{p\alpha} d^{}_a $ has two terms, corresponding to electrons hopping from the bath to the impurity and from the impurity back to the bath, only even perturbation orders appear in the expansion of $Z$. 
Furthermore, at perturbation order $2n$, only the $(2n)!/(n!)^2$ terms corresponding to $n$ creation operators $d^\dagger$ and $n$ annihilation operators $d$ contribute. We therefore write the partition function as a sum over configurations $\{\tau_1,\ldots, \tau_n;\tau_1',\ldots,\tau_n'\}$ that are collections of imaginary-time points corresponding to these $n$ annihilation and $n$ creation operators:
\begin{eqnarray}
&&Z=\sum_{n=0}^\infty \int_0^\beta d\tau_1\cdots \int_{\tau_{n-1}}^\beta d\tau_n 
\int_0^\beta d\tau_1'\cdots \int_{\tau_{n-1}'}^\beta d\tau_n'\nonumber\\
&& \times {\rm Tr}
\Big[ e^{-\beta H_1}{\cal T} H^d_\mathrm{hyb}(\tau_n) H^{d^\dagger}_\mathrm{hyb}(\tau_n') \cdots H^d_\mathrm{hyb}(\tau_1) H^{d^\dagger}_\mathrm{hyb}(\tau_1')\Big].\nonumber
\label{Z_interaction_picture_strong}
\end{eqnarray}

Introducing the $\beta$-antiperiodic hybridization function $\Delta$, which in the %time-domain reads  
Matsubara-frequency representation reads
\begin{eqnarray}
&&\Delta_{ab}(i\omega_n)=\sum_{p,\alpha} \frac{(V^{a\alpha}_{p})^*(V^{b \alpha}_p) }{i\omega_n-\varepsilon_{p\alpha}},
%&&\Delta_{ab}(\tau)=\sum_{p,\alpha} \frac{(V^{a\alpha}_{p})^*(V^{b \alpha}_p) }{e^{\varepsilon_{p\alpha}\beta}+1}
\nonumber
\end{eqnarray}
we can explicitly evaluate the trace over the bath states to find $Z_{\rm bath} \det M^{-1}$, where $M^{-1}$ is an $(n \times n)$ matrix with elements
\begin{equation}
[M^{-1}]_{ij}=\Delta_{a_i'a_j}({\tau'_i}^{a_i'}-\tau_j^{a_j}).
\label{M}
\end{equation}
In the hybridization expansion approach, the configuration space thus consists of all sequences  
$C=\{\tau_1^{a_1}, \ldots,\tau_{n}^{a_n}; \tau_1'^{a_1'}, \ldots,\tau_{n}'^{a'_n} \}$
of $n$ creation and $n$ annihilation operators ($n=0,1,\ldots$), %, see illustration in Fig.~\ref{fig:conf}(a). The 
and the weight of such a configuration is
\begin{eqnarray}
w_C&=&Z_\mathrm{bath}
{\rm Tr}_d \Big[e^{-\beta H_\mathrm{loc}} {\cal T} 
d_{a_n}(\tau_{n}^{a_n})d^\dagger_{a_n'}(\tau_{n}'^{a_n'})
\cdots\nonumber\\
&&\cdots
d_{a_1}(\tau_1^{a_1})d^\dagger_{a_1'}(\tau_1'^{a_1'}) 
\Big]\det M^{-1} (d\tau)^{2n}.\nonumber
\label{weight_strong}
\end{eqnarray}
The trace factor represents the contribution of the impurity, which fluctuates between different quantum states as electrons hop in and out, while the determinant sums up all bath evolutions which are compatible with the given sequence of transitions.

%The hybridization functions can in general be off-diagonal and complex ($\Delta_{ab}\ne 0$ for $a\ne b$), which then leads to complex configuration weights $w_C$, and to a sign problem (or more properly phase problem) in the Monte Carlo sampling.
The hybridization functions can in general be off-diagonal ($\Delta_{ab}\ne 0$ for $a\ne b$) and complex 
and the trace factor also can be complex in the presence of SOC. This typically leads to complex configuration weights $w_C$, and to a sign problem (or more properly phase problem) in the Monte Carlo sampling.
We can implement a Monte Carlo sampling using the positive distribution of weights $|w_C|$, and shift the phase to the observable. As shown in Eq.~(\ref{phase}), the expectation values of observables are then obtained by the ratio of the phase weighted measurement and the average value of the phase. 
Note that while $w_C/|w_C|$ can be a complex number, 
the expectation value $\langle e^{i\varphi}\rangle$, which is usually denoted by ``average sign'', is always real because the partition function is real.

An open source hybridization expansion continuous-time Monte Carlo impurity solver, which handles off-diagonal and complex hybridizations, and thus is suitable for the LDA+DMFT study of pyrochlore iridates, has recently been published in Ref.~\cite{Shinaoka:2017cpa}. 

\subsubsection{Sign problem and choice of local basis }
The sign problem strongly depends on the local basis in which the hybridization expansion is performed. By transforming from orbitals $\{\alpha\}$ to new orbitals $\{\alpha'\}$ we change the hybridization functions, and hence the diagrams which are sampled. %Often it is a good strategy to work in a basis in which the hybridization matrix is diagonal or close to diagonal. 
Typically, the sign problem is severe if the off-diagonal components of the hybridization function are large. 

In some previous studies of pyrochlore iridates, the quantum impurity models for $5d$ electrons have been simplified to avoid a severe sign problem by omitting off-diagonal hybridization functions and some interaction terms in the $\jeff$ basis. However, since pyrochlore iridates have large inter-band hybridizations, one should avoid such approximations. The exact treatment also does not assume the quantization axes of spin and orbital.
The sign problem in calculations with full hybridization matrix can to some extent be reduced by rotating the single-particle basis of the hybridization function.

Specifically, for the calculations in Fig.~\ref{fig:iridate-pd}, we solved the quantum impurity problem by Monte Carlo sampling in a basis which diagonalizes the local SOC in the $t_\mathrm{2g}$ manifold.
This eigenbasis was constructed by numerically diagonalizing the non-interacting Hamiltonian on an Ir atom.
Although the basis vectors correspond to the three doublets shown in the diagram of Fig.~\ref{fig:CF}(b), the phase degrees of freedom in each doublet were numerically fixed.
Thus, the basis functions are eigenvectors of $(\hat{j}_\mathrm{eff}^{111})^2$ but not $\hat{j}_\mathrm{eff}^{111}$.\footnote{We did not observe any significant dependence of the average sign on the choice of the phase degrees of freedom within the doublets.}
In the eigenbasis,
we observed substantial off-site hybridization functions between the lowest and the highest doublets, leading to a sign problem in the Monte Carlo sampling.
We took into account all the off-diagonal hybridization functions using the impurity solver of Ref.~\cite{Shinaoka:2017cpa}.

\begin{figure}[h]
	\centering\includegraphics[width=.4\textwidth,clip]{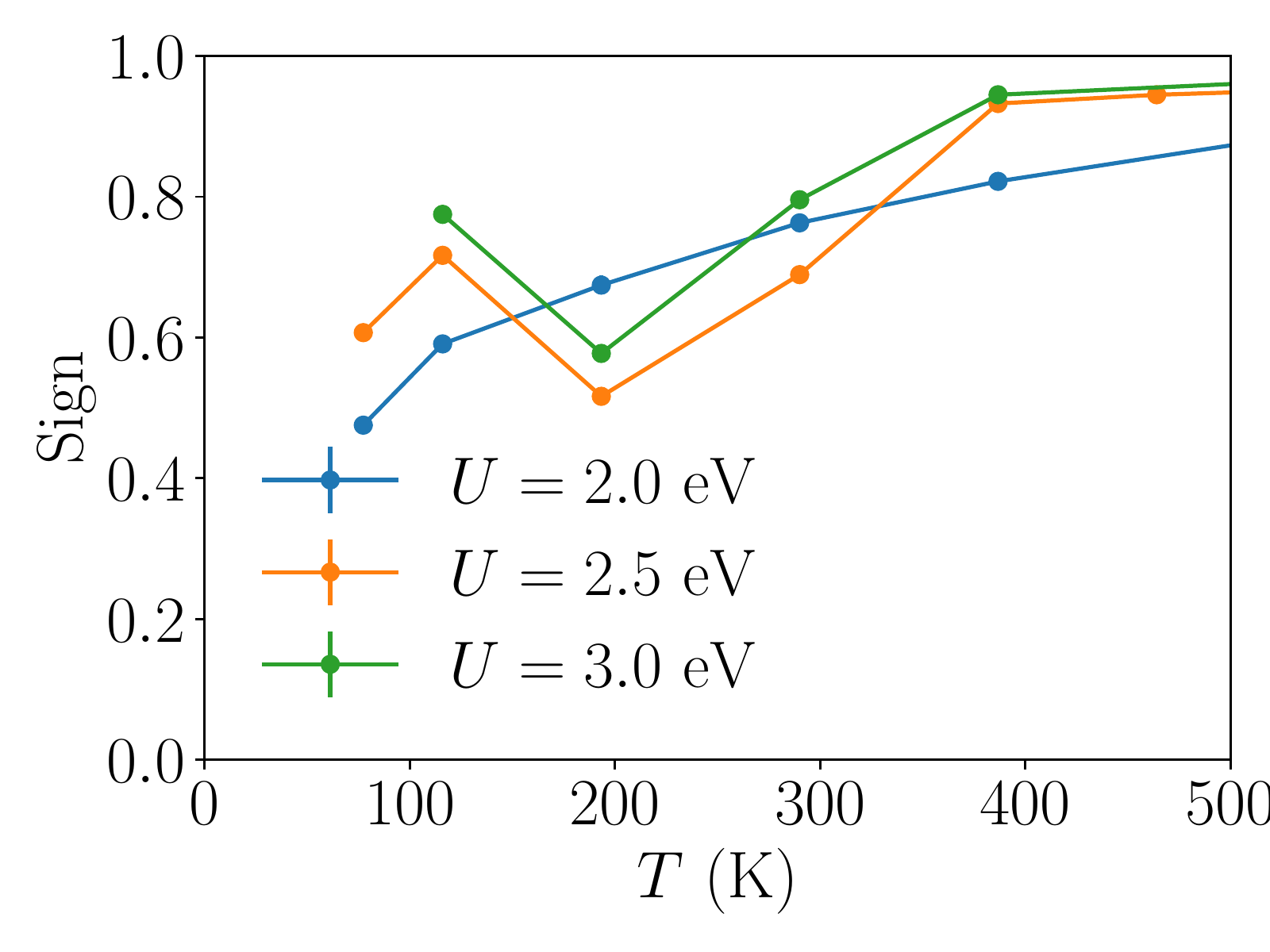}
	\caption{
		(Color online) Average sign encountered in solving the DMFT quantum impurity problem for pyrochlore iridates. The average sign remains above 0.4 down to 77 K for $U$ = 2.5 and 3 eV.
	}
	\label{fig:iridate-sign}
\end{figure}

Figure~\ref{fig:iridate-sign} shows the average sign encountered in solving the quantum impurity problems.
The calculations were done for $J_{\rm H}=0.1U$ down to $\beta=150$ (1/eV), corresponding to 77 K.
The average sign remains sufficiently large in the eigenbasis that calculations can be performed with modest
computational resources. 
Interestingly, the temperature dependence of the sign is not monotonic.
At large $U$, the average sign shows a local minimum around 200 K, recovers, and then decreases again below 100 K.
Although this basis may not be optimal in terms of average sign,
it turns out to be good enough for the purpose of mapping out the phase diagram of Y$_2$Ir$_2$O$_7$.

%\clearpage
%\section{Other related compounds}

\section{Summary and future perspectives}\label{sec:summary}

In this article, we have reviewed our first-principles studies on $4d$ and $5d$ pyrochlore oxides, where spin-orbital physics plays a substantial role.
We showed that in the insulating Mo pyrochlores, the interplay of spin and orbital under the SOC is relevant to the magnetic properties such as the puzzling spin-glass behavior.
The SOC also plays an important role in stabilizing the non-collinear magnetic order and determining the metal-insulator transition in the Os compound Cd$_2$Os$_2$O$_7$.
For the Ir pyrochlores, LDA+DMFT calculations within a three-orbital description qualitatively reproduce the experimental phase diagram, and we emphasized the importance of the multi-orbital nature.
We also reviewed the theoretical frameworks used in these studies.

Let us briefly discuss the future direction of first-principles studies on pyrochlore oxides and related compounds.
In section~\ref{sec:iridate-pd}, the on-site $U$ was considered to be a control parameter for the metal-insulator transition at half filling in pyrochlore iridates.
However, is this \textit{assumption} really correct?
Which microscopic parameters change as we go through the metal-insulator transition in Fig.~\ref{fig:iridate-pd-exp}?
This is an open issue which touches on fundamental properties of the pyrochlore iridates.
In the supplemental material of Ref.~\cite{Zhang:2017ib}, using the LDA+DMFT calculations,
Zhang \textit{et al.} pointed out the important role of the internal oxygen coordinate $x$ in determining the ground state of the pyrochlore iridates.
On the other hand, a recent cluster DMFT study of single-orbital tight-binding models indicated the importance of intersite magnetic correlations
in the qualitative description of the $A$-site dependence~\cite{Wang:2017ic}.
These effects are not captured by single-site(atom) LDA+DMFT calculations.
Thus, a comprehensive understanding of the metal-insulator transition is still missing. 

Another interesting topic for $5d$ pyrochlores is the design of non-trivial quantum phases at surfaces, magnetic domain boundaries, or  in superlattice structures.
Recently, the existence of metallic states at magnetic domain boundaries has been confirmed experimentally for Cd$_2$Os$_2$O$_7$~\cite{Hirose:2017gl}, Eu$_2$Ir$_2$O$_7$~\cite{Fujita:2015cab,Fujita:2016hw}, and Nd$_2$Ir$_2$O$_7$~\cite{Ma:2015du}.
On the other hand, it has been theoretically shown that such magnetic boundaries can host metallic states with non-trivial topological properties~\cite{Yamaji:2014fa,2016PhRvB..93s5146Y}.
Furthermore, there are interesting theoretical proposals for topological phenomena in thin films of pyrochlore iridates~\cite{Hu:2015cc,Yang:2014cx}.
Designing such non-trivial quantum phases by first-principles calculations may be an interesting topic.

Recently, the 5$d$ pyrochlore Cd$_2$Re$_2$O$_7$ with a $5d^2$ configuration has attracted much attention.
This compound is unique in the sense that among the ($\alpha$-)pyrochlores,
it is the only one which shows superconductivity (at $T\simeq 1~K$)~\cite{2001JPCM...13L.785S,2002PhyC..378...43S,Hanawa:2001bh}, and hosts various unidentified (likely) electronic phases.
At ambient pressure, the compound exhibits two successive structural transitions at $T\simeq 200$ K and $T\simeq 120$ K~\cite{Hanawa:2001bh,Yamaura:2002fk,Castellan2002,Weller:2004eh,Sergienko:2004de,Kendziora:2005cc,Petersen:2006ih,Harter2017,Harter:2018ib,Hiroi:2018hm}.
Despite drastic changes in the electronic properties below these transition temperatures~\cite{Vyaselev:2002bz,Hiroi:2003fa},
only tiny structural changes were observed, indicating that their origins are electronic.
%The inversion symmetry of the crystal structure is broken below $T\simeq 200$ K 
%these transition temperatures~\cite{Hanawa:2001bh}.
%
A recent experimental study has reported the emergence of multipolar phases below these transition temperatures~\cite{Harter2017}.
It was also suggested that this compound is a prototype of the spin-orbit-coupled metal recently proposed in Ref.~\cite{Fu:2015cl}.
The nature of these transitions is however under debate.
%More interestingly, 
Remarkably, 
at high pressures, the compound exhibits an even richer structural phase diagram~\cite{Yamaura2017}.

There obviously remains a broad spectrum of interesting and unsolved problems, especially in the study of $5d$ pyrochlores. Theoretical investigations of these compounds will play an important role in clarifying the relevant physical processes, and these applications will mark a frontier of modern first-principles simulations of correlated systems.

\ack{
HS and PW acknowledge support from the DFG via FOR 1346, from SNF Grant No. 200021E-149122, ERC Advanced Grant SIMCOFE and NCCR MARVEL. 
}
\hspace{1cm}
\mbox{}

\bibliographystyle{unsrt}
%\bibliography{ref.bib,ref_ym.bib,ref_miyake.bib,ref_aux.bib}

%\subsection{Appendices}
% 
%\subsection{Something}

\end{document}